\documentclass[twocolumn]{aastex62}
\usepackage{graphicx} 
\usepackage{booktabs}
\usepackage{xcolor}
\usepackage{url}
\usepackage{longtable} 
\usepackage{times}
\usepackage{float}
\usepackage{wrapfig}
\usepackage{rotating}
\usepackage{amssymb}
\usepackage{dejavu}
\usepackage{colortbl,booktabs,array,tabularx}
\usepackage{lipsum,graphicx,float}
\usepackage[caption=false]{subfig}
\usepackage{graphicx}
\usepackage{float}
\usepackage{scrextend}
\usepackage{hyperref}
\usepackage{textcomp}
\usepackage{multirow}

\newcommand\aastex{AAS\TeX}

\newcommand{\mum}{\textmu m}
\accepted{by The Astrophysical Jornal, 2025 August 16}

\shorttitle{\aastex\ Diffuse ISM Dust}
\shortauthors{Pendleton et al.}

\begin{document}

\title{A Tale of Two Sightlines: Comparison of Hydrocarbon Dust Absorption Bands toward Cygnus OB2-12 and the Galactic Center}
\author[0000-0001-8102-2903]{Yvonne J. Pendleton}
\affiliation{Department of Physics, University of Central Florida, Orlando, FL 32816, USA}

\correspondingauthor{Yvonne Pendleton}
\email{pendletonyvonne@gmail.com}

\author[0000-0003-2824-3875]{T. R. Geballe}
\affiliation{Gemini Observatory/NSF's NOIRLab, 670 N. A'ohoku Pl., Hilo, HI 96720, USA}

\author[0000-0002-1437-4463]{Laurie E. U. Chu}
\affiliation{NSF NOIRLab,950 N. Cherry Ave., Tucson, AZ, 85719, USA}

\author[0000-0001-9462-5543]{Marjorie Decleir}
\altaffiliation{ESA Research Fellow}
\affiliation{European Space Agency (ESA), ESA Office, Space Telescope Science Institute, 3700 San Martin Drive, Baltimore, MD, 21218, USA}
  
\author[0000-0001-5340-6774]{Karl D.\ Gordon}
\affiliation{Space Telescope Science Institute, 3700 San Martin Drive, Baltimore, MD, 21218, USA}
\affiliation{Sterrenkundig Observatorium, Universiteit Gent, Krijgslaan 281 S9, B-9000 Gent, Belgium}

\author[0000-0003-0306-0028]{A.G.G.M. Tielens}
\affiliation{Astronomy Department,  University of Maryland, 4296 Stadium Drive, College Park, MD 20742-2421}

\author[0000-0002-6049-4079]
%0000-0001-6035-3869]
{Louis J. Allamandola}
\affiliation{NASA Ames Associate-retired, NASA Ames Research Center, M/S 245-6, Moffett Field, CA 94035, USA}

\author[0000-0003-4757-2500]{Jeroen Bouwman}
\affiliation{MPI for Astronomy, Koenigstuhl 17, 69117 Heidelberg, Germany}

\author[0000-0003-2029-1549]
%0000-0003-1120-5178]
{J. E. Chiar}
\affiliation{Physical Science Department, Diablo Valley College, 321 Golf Club Road, Pleasant Hill, CA 94523, USA}

\author[0000-0002-6528-3836]{Curtis Dewitt}
\affiliation{Space Science Institute, Boulder, CO 80301, USA}

\author[0000-0002-2449-0214]{Burcu Gunay}
\affiliation{Armagh Observatory and Planetarium, Armagh, NI, UK}

\author[0000-0002-1493-3000X]{Thomas Henning}
\affiliation{MPI for Astronomy, Koenigstuhl 17, 69117 Heidelberg, Germany}

\author[0000-0001-9525-895X]{Vito Mennella}
\affiliation{INAF Osservatorio Astronomico di Capodimonte, Salita Moiariello 16, 80131 Napoli Italy}

\author[0000-0002-9122-491X]{M. E. Palumbo}
\affiliation{INAF - Osservatorio Astrofisico di Catania, via Santa Sofia 78, 95123 Catania, Italy}

\author[0000-0002-3699-7477]
%0000-0002-9518-4679]
{Alexey Potapov}
\affiliation{Institute of Geosciences, Friedrich Schiller University Jena, Jena, Germany}

\author[0000-0002-3936-2469] 
%0000-0001-5400-1461]
{Maisie Rashman}
\affiliation{The Open University, Walton Hall, Milton Keynes MK7, 6AA, United Kingdom}

\author[0000-0002-8163-8852]{Sascha Zeegers}
\altaffiliation{ESA Research Fellow}
\affiliation{European Space Agency (ESA), European Research and Technology Centre (ESTEC), Keplerlaan 1, 2201 AZ Noordwijk, The Netherlands}

\begin{abstract}

Infrared spectra of hydrocarbon dust absorption bands toward the bright hypergiant Cygnus OB2-12 are compared to published spectra of the Quintuplet Cluster, a sightline to the Galactic center. The Cyg OB2-12 data include a new ground-based 2.86$-$3.70~\mum\ spectrum and a previously published, but here further analyzed, spectrum of the 5.50--7.34~\mum\ region. Higher spectral resolution data for the  Cyg OB2-12 sightline in the 3 \mum\ region allows a detailed comparison of the 3.4~\mum\ aliphatic bands to those observed toward the Quintuplet. Despite differences in interstellar environments along each sightline, strong similarities are observed  in the central wavelengths and relative strengths for bands at $\sim$ 3.3, 3.4, 5.85, 6.2, and 6.85~\mum. Analysis of these bands, produced by aromatic, aliphatic, olefinic, hydrogenated, and oxygenated components, shows that carbonaceous dust is a significant component of the diffuse interstellar medium, second in abundance only to silicates, and is primarily aromatic in nature. The grains producing these bands likely consist of large aromatic carbon cores with thin aliphatic mantles composed of hydrogenated amorphous carbon (HAC). Laboratory analog spectra reproduce the observed aliphatic absorption bands well, supporting the presence of such mantles. We present evidence that the carriers of both the 3.4 \mum\ aliphatic and the 3.3~\mum\ aromatic bands reside exclusively in the diffuse ISM, and that the 3.3~\mum\ bands observed in the diffuse ISM differ from the 3.25 \mum\ band seen in dense clouds, implying chemically distinct carriers.


\end{abstract}

\section{Introduction}
\label{sec:intro}

The composition of interstellar dust is the cumulative result of numerous production and destruction processes whose rates vary considerably with the environment, particularly between low-density diffuse clouds and dense molecular clouds \citep[e.g.,][]{dorschner1995, jones2012a, jones2012b, tielens2005physics, chiar2013structure}. Dust is produced in the expanding shells of oxygen-rich and carbon-rich evolved stars and supernovae, and is transported into the diffuse interstellar medium (ISM) by outflowing winds. Observations and theoretical models have shown that dust can also form and be substantially modified within the ISM itself, particularly through accretion and processing in low-density environments \citep[e.g.,][]{dorschner1995,jones2011dust}. Some components, such as aliphatic hydrocarbons, may be formed directly in the diffuse ISM \citep[e.g.,][]{mennella2002ch,jones2011dust,jones2012a,jones2012b}. In diffuse interstellar clouds, H$_2$ can be formed efficiently on grain surfaces \citep[e.g.,][]{hollenbachSalpeter1971,jura1974}.

Thereafter, dust grains that survive the rigors (and modifications) of the diffuse interstellar environment can be swept into denser clouds, where they provide surfaces on which simple species can react to form more complex molecules, resulting in the formation of icy grain mantles \citep{tielensHagen1982,hama-icechemistryreview,Boogert15}. 
Potential pathways for creating more complex materials from the initial simple species formed by grain surface reactions have also been identified in the laboratory  \citep[e.g.,][]{tielens2013molecular,potapov2021dust, theule2013thermal}. 
Theoretical modeling studies, such as those by \citet{garrod2008, garrod2013}, have complemented laboratory results by simulating grain-surface and bulk ice chemistry under astrophysical conditions, including the formation and evolution of complex organic molecules.
Species as complex as methanol (CH$_{3}$OH), ethanol (CH$_{3}$CH$_{2}$OH), acetaldehyde (CH$_{3}$CHO), and methyl formate (CH$_{3}$OCHO) ice have been observed in dense molecular cores in which star formation is either imminent or is occurring \citep[e.g.,][]{Boogert15, Chu2020, chen2024joys+, rocha2024jwst}.

Infrared spectroscopy is ideally suited for identifying and quantifying the composition and structure of interstellar dust, as many diagnostic bands occur at wavelengths between 2.5 \mum\ and 25~\mum. However, while the study of dust composition in dense clouds has flourished for decades, the same cannot be said for the study of dust in the diffuse ISM. This difference is mainly due to the lower extinctions (and therefore lower column densities of intervening dust) in diffuse clouds toward sources with sufficient brightness to enable sensitive spectral profiles of the weak absorption bands to be obtained. Recent and forthcoming spectroscopy from the James Webb Space Telescope (JWST) promises to drastically change this disparity; however, this paper presents new and re-analyzed observations of a very important diffuse ISM sightline that cannot be studied by JWST due to the extreme brightness of the background star, Cygnus OB2-12 (hereafter Cyg OB2-12).
  
Two vastly different sightlines, the Galactic center (GC) Quintuplet star cluster (\cite{chiar2013structure}; $A_{V}$ $\sim$ 29 mag) and Cyg OB2-12 (\cite{whittet1997infrared}; $A_{V}$ $\sim$ 10 mag), both substantially extinguished by intervening dust, have been studied through infrared spectroscopy using  ground-based, airborne, and space telescopes over the decades, improving our understanding of dust commensurate with improvements in telescope and instrument capabilities. Although the two sightlines are complicated in different ways, by intervening dense cloud material in the case of the GC and in both cases, by the very different environments local to their background sources, both sightlines show a strong, broad absorption band at 3.4~\mum, attributed to aliphatic hydrocarbons \citep[][and others]{Soifer1976, Willner1979, Persi82, Roche85, adamson1990, sandford1991interstellar, pendleton1994near, whittet1997infrared, pendleton2002organic, chiar2002infrared, chiar2013structure, dartois2007carbonaceous}. The primary impetus for our study is the similarity of this common broad absorption band.

While sightlines with lower extinction have primarily been used to develop infrared extinction curves \citep{gordon2021milky,decleir2022spex}, the Cyg OB2-12 sightline was one of the earliest sightlines utilized  \citep{rieke85}.  A more complete understanding of the dust composition contributing to the extinction along this sightline is necessary if Cyg OB2-12 is to be used in modern day extinction curve studies.  Here we provide a next step in understanding the chemical composition of the intervening dust towards Cyg OB2-12  through infrared spectral comparisons of the GC and Cyg OB2-12 sightlines.
It is generally expected that a roughly linear relationship should exist between the amount of intervening dust in the diffuse ISM and the strength of the 3.4~\mum\ band. Still, the tightness of any such correlation remains unclear.  Measurements of the 3.4~\mum\ band by \cite{sandford1991interstellar} and \cite{pendleton1994near} on sightlines other than toward the GC and Cyg OB2-12 are consistent, albeit with considerable uncertainty, with a linear relationship for $A_{V}$ $\leq$ 13 mag. However, for a different set of sightlines with a similar range of extinctions as \cite{pendleton1994near}, \cite{rawlings2003infrared} found a significantly weaker band strength per unit $A_{V}$. Further progress on the relationship between $A_{V}$ and the strength of the 3.4~\mum\ absorption band awaits JWST observations along multiple sightlines.

In addition to the broad 3.4~\mum\ absorption band, a much weaker absorption band at $\sim$ 3.2$-$3.3~\mum\  has also been detected in the spectra of sightlines towards both Cyg OB2-12 and GC sources \citep{Hensley20,chiar2013structure}, which they and we refer to as the 3.3~\mum\ band. The band has also been observed and analyzed in detail on an unrelated GC sightline \citep{geballe2021interstellar,bernstein2024analysis}. A somewhat similar absorption feature has been observed toward several deeply embedded young stellar objects (YSO's)   \citep{sellgren94, sellgren95, brooke1996study,  brooke1999}, which those authors refer to as the 3.25~\mum\ band. Given the weakness of these features and the consequent uncertainty regarding their spectral profiles, it has been unclear how closely the feature observed toward YSOs in dense cores matches those seen toward Cyg OB2-12 and the GC, although both can be confidently ascribed to the CH stretch in aromatic hydrocarbon species. In this paper, we compare the two bands in detail and examine their relationship.

Additional key constraints on the chemical nature of the diffuse dust toward Cyg OB2-12 and the GC can be obtained from their spectra in the 5.5$-$7.0~\mum\ region. This interval contains the C-C stretch bands from olefins and aromatics at 6.19 and 6.25~\mum, respectively,the C-H scissoring mode from aliphatics (due to bending within CH$_{2}$ groups) at 6.85~\mum\ \citep[e.g.][]{allamandola1984absorption, tielens1996infrared}, along with C=O stretch bands due to carbonyls and quinones at 5.85 and 6.05~\mum, respectively (see Section \ref{sec:sightlines and chemistry} for details related to chemical terms used in this paper). 

\citet{chiar2013structure} conducted a comprehensive study of most of the above bands toward the GC Quintuplet Cluster. They concluded that interstellar  carbon dust along  that sightline consists of grains with primarily aromatic carbon cores covered with thin mantles of hydrogenated amorphous carbon (HAC). They also found that Galactic hydrocarbon dust is highly aromatic, with a low overall H/C ratio. 

In this paper, we compare their conclusions to those we have reached from analysis of these bands on the Cyg OB2-12 sightline.  We use a new near-infrared (hereafter near-IR) spectrum of Cyg OB2-12, obtained at the \textit{NASA Infrared Telescope Facility} (IRTF), covering the 3.3~\mum\ and 3.4~\mum\ hydrocarbon features at similar sensitivity but significantly higher spectral resolution than previously reported. We also include a new presentation and analysis of the 5.50$-$7.34~\mum\ portion of the mid-infrared (hereafter mid-IR) spectrum toward Cyg OB2-12 obtained by the \textit{Spitzer Space Telescope}. This portion of the spectrum was previously presented by \cite{Hensley20} and \cite{potapov2021dust}, but was not shown by them in detail and was not subjected to the type of analysis performed here.

Recent observations by the JWST have begun to extend our understanding of hydrocarbon and silicate absorption features in the diffuse ISM. The Measuring Extinction and Abundances of Dust (MEAD) project has reported tentative detections of the 3.4 and 6.2~\mum\ features in sightlines with visual extinctions as low as $A_V \sim~$2.5 mag \citep{Decleir2025}, suggesting that carbonaceous carriers persist even in relatively low-extinction diffuse environments. Similarly, early results from the Webb Investigation of Silicates, Carbons, and Ices (WISCI) project demonstrate that these features are robustly detected along multiple diffuse sightlines observed at high signal-to-noise with JWST \citep{Zeegers2025}. Although JWST cannot observe the bright hypergiant Cyg OB2-12 directly, the presence of comparable hydrocarbon absorption bands in its spectrum allows us to connect earlier high-extinction studies with these new JWST results. This study provides a bridge between these regimes, offering a benchmark for interpreting carbonaceous absorption features across a broad range of diffuse ISM conditions.

This paper is organized as follows: Section \ref{sec:observations} contains descriptions of the observations, data reduction, and the fitting of band profiles by one or multiple Gaussians. Section \ref{sec:Analysis} presents our analysis and the basic results that can be directly inferred from the observations.   Section \ref{sec:Disc} reviews our results in the context of the formation, evolution, and structure of interstellar hydrocarbon dust, including a comparison of the observations to three laboratory analogs. Section \ref{sec:Summary} presents our conclusions from this work.

\section{Observations, Data Reduction, and Profile Fitting}
\label{sec:observations}

\subsection{Cyg OB2-12 Near-IR Observations}
\label{sec: NIR obs}

A 2.0$-$5.3~\mum\ spectrum of Cyg OB2-12 was obtained at the NASA  Infrared Telescope Facility (IRTF) on UT 2022 July 12 using the facility instrument SpeX \citep{rayner2003spex} in its long-wavelength, cross-dispersed mode (Co-PIs: Pendleton and Chu). Five sets of Cyg OB2-12 spectra were obtained, interspersed with four sets of spectra of HD192538  (A0V), which served as the telluric standard star. All spectra were obtained in the standard nod-along-slit mode, with a nod distance of 7\farcs5 and a slit width of 0\farcs3. The resolving power R (=$\lambda$/$\Delta\lambda$) was approximately 2500, which corresponds to a wavelength resolution of 0.0013~\mum\ at 3.3~\mum. The total exposure time on Cyg OB2-12 was 22.2 minutes.

The SpeX data reduction tool \citep{cushing2004} provided for each of the five separate observations of Cyg OB2-12 and four separate observations of the standard star, wavelength-calibrated spectra of the signal, in several grating orders across the 2.0$-$5.3~\mum\ interval. For the purposes of this work, we utilized only orders 2 and 3 which cover the wavelength range of interest, $\sim$ 3$-$4~\mum .  We further restricted our additional data reduction in order 2 to 3.10$-$3.70~\mum\ and in order 3 to 2.86$-$3.15~\mum.  Using routines in Starlink Figaro \citep{Currie14}, we resampled the spectra of both stars in 0.0005 \mum\ bins, in order to put all of them on the same wavelength scale. We then ratioed each of the five Cyg OB2-12 spectra by one standard star spectrum and flux-calibrated each ratioed spectrum, using standard procedures.

For each of the two SpeX orders, the five individual flux-calibrated spectra of Cyg OB2-12 were coadded. Each individual reduced spectrum as well as their coadd showed residual telluric emission features in both wavelength segments, due to the mismatch in airmass between Cyg OB2-12 and the telluric standard, as well as artificial \ion{H}{1} emission lines, created by photospheric absorption lines in the spectrum of the telluric standard. The latter can be removed by standard techniques (see below). The former can severely compromise the interpretation of the spectrum, especially in the 2.86$-$3.50~\mum\ interval, which contains many very strong telluric absorption lines of H$_{2}$O and CH$_{4}$. 

To approximately remove these residuals, we used the technique described in detail by \cite{bernstein2024analysis} and summarized in this and the following paragraph for this particular usage. (1)  We converted  the coadded F$_{\lambda}$ spectrum of Cyg OB2-12 to an optical depth spectrum as follows. (1a) To flatten the Cyg OB2-12 spectrum, we calculated and plotted the ratio of the  F$_{\lambda}$ spectra of the source to that of a 13,700 K blackbody, the effective temperature of Cyg OB2-12 \citep{clark2012nature}. (1b)  We then approximated the flattened spectrum by a linear continuum connecting wavelengths near the short and long wavelength extremities that are both devoid of stellar spectral features and correspond to wavelengths of nearly 100\%\ atmospheric transmission, based on ATRAN model atmospheric transmission spectra \citep{Lord1992}. (1c) After dividing by the continuum, we converted the result to an optical depth spectrum (which still contained the residual telluric features and the artificial \ion{H}{1} emission lines).  (2) We applied steps similar to the above to the coadded spectra of the telluric standard, again approximating the continuum by linear segments between wavelengths of $\sim$100 percent atmospheric transmission near 3.10~\mum\ and 3.70~\mum\ for order 2 and near 2.86~\mum\ and 3.15~\mum\ for order 3, and then dividing the observed spectrum by the continuum. We then converted the resulting ratioed spectra to atmospheric optical depth spectra (which also contained \ion{H}{1}  absorption lines of the star). 

(3) After a slight smoothing of the Cyg OB2-12 optical depth spectra to match the slightly lower resolution of the spectrum of the standard star, we scaled the atmospheric optical depth spectra (by eye) to match the residual telluric features in the Cyg OB2-12 optical depth spectra \citep[see Figure 3 of][for an example]{bernstein2024analysis}. (4) The scaled atmospheric optical depth spectra were subtracted from the Cyg OB2-12 optical depth spectra to remove the residuals and produce an optical depth spectrum of the Cyg OB2-12 sightline for each order. (5) The final 2.86--3.15 \mum\ and 3.10--3.70 \mum\ spectra were adjoined, after re-normalizing one of them slightly so that their average values in the overlapping 3.10$-$3.15~\mum\ interval were identical. (6) Finally we added to the adjoined spectrum a continuum-normalized model spectrum of an A0V star, converted to optical depth, in order to remove the false \ion{H}{1} emission features produced by ratioing Cyg OB2-12 by the A0V telluric standard. 

\begin{figure*}[!htb]
\epsscale{1.1}
\includegraphics[trim={0 0 0 0 0},clip,width=1.0\textwidth]{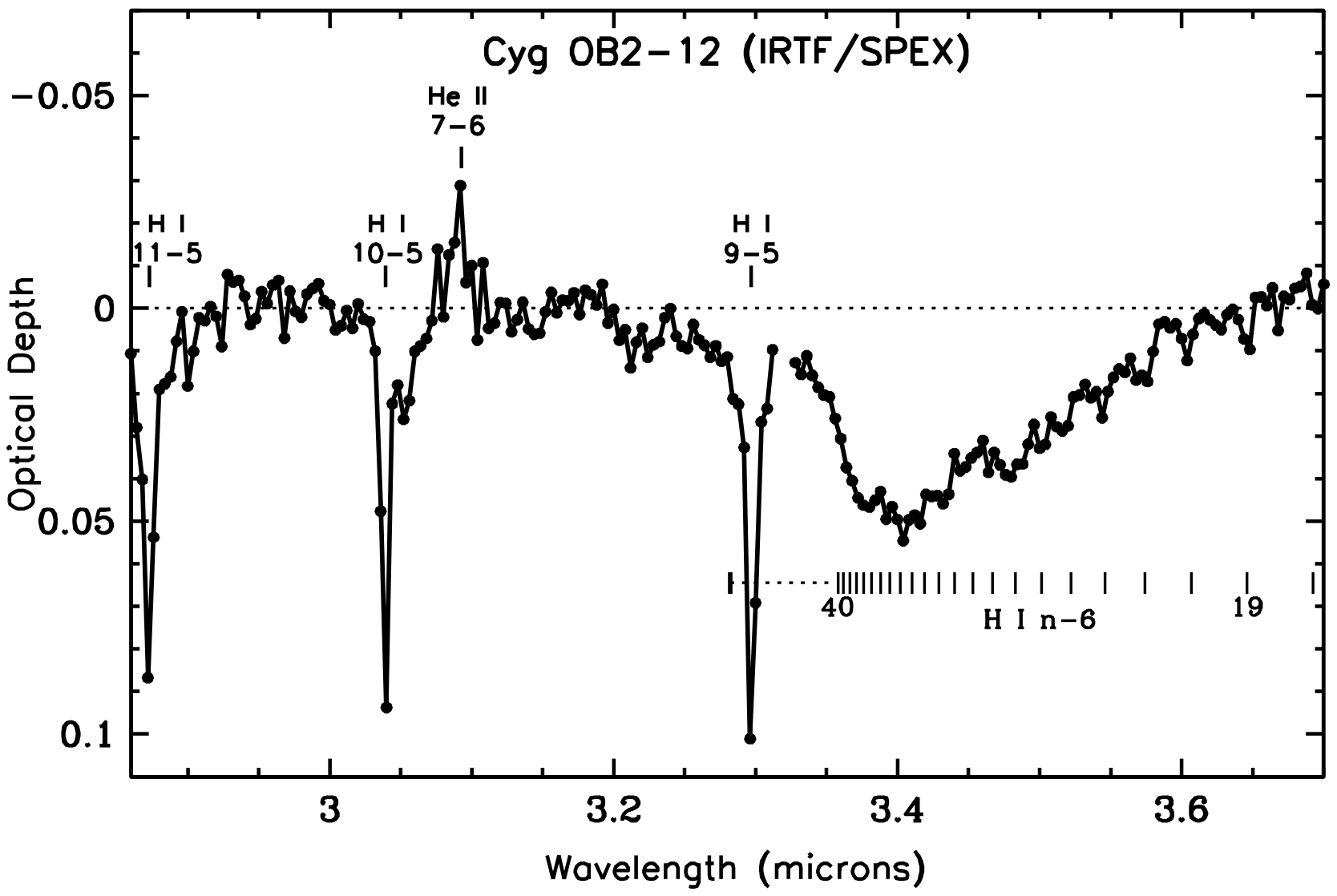}
\caption{The 2.86$-$3.70~\mum\ optical depth spectrum of Cyg OB2-12 obtained with SpeX on the NASA IRTF. The spectrum is displayed at a resolution of 0.004~\mum. Telluric line residuals were removed as described in \S~\ref{sec: NIR obs}. Three \ion{H}{1}  Pfund series ({\it n}$-$5) absorption lines and a weak \ion{He}{2} emission line are labeled. The wavelengths of the  \ion{H}{1}  Humphreys series ({\it n}$-$6) lines and the series limit at 3.283~\mum\ are indicated by short vertical lines. Data points in the interval  3.31$-$3.32~\mum\ are spurious  due to incomplete correction of the nearly totally opaque telluric absorption by closely spaced lines of the methane $\nu_{3}$ Q-branch, and are not shown. The noise level varies with wavelength but can be estimated from the point to point fluctuations in the signal in regions devoid of absorption features, such as 2.90$-$3.00~\mum\ and 3.12$-$3.20~\mum.}
\label{fig:fig1}
\end{figure*}

Different scaling factors were employed in the 2.86$-$3.50~\mum\ interval, where the atmospheric transmission spectrum contains many optically thick lines, and the 3.50$-$3.70~\mum\ interval, where almost all of the telluric lines are optically thin \citep[see Appendix 1 of][]{bernstein2024analysis}. A finer division of the spectrum into sub-spectra and the application of  different scaling factors to each division might slightly improve the result, but considering the signal-to-noise ratio of the spectrum, we find the current (coarser) approach to be adequate.

Figure \ref{fig:fig1} shows the 2.86$-$3.70~\mum\ optical depth spectrum of Cyg OB2-12 derived from the IRTF data, which represents the best quality spectrum available from ground-based observations in this wavelength interval.  The spectrum was obtained at a resolution of $\sim$ 0.0013~\mum, but is shown here in bins of 0.004~\mum, which corresponds to a resolving power, R $\sim$ 800. It has a similar signal-to-noise ratio as the spectrum presented by \citet {Hensley20} at lower R (200), which was obtained by the Short-Wavelength Spectrometer (SWS) of the \textit{Infrared Space Observatory} (ISO). In addition to the relatively strong 3.4~\mum\ band, a weak absorption feature, extending from $\sim$3.21~\mum\ to somewhat beyond 3.30~\mum\ is clearly present, with the \ion{H}{1} 9$-$5 absorption line superimposed on part of it. We henceforth refer to this band as the 3.3~\mum\ band. The profile of the 3.4~\mum\ absorption is similar to that reported on other diffuse cloud sightlines \citep{sandford1991interstellar,pendleton1994near,rawlings2003infrared}. The low signal-to-noise ratio of the weak 3.3~\mum\ absorption band makes it difficult to characterize its profile.   

Three Pfund series absorption lines of \ion{H}{1}, {\it n} = 9$-$5, 10$-$5, and 11$-$5  at 3.297, 3.039, and 2.873~\mum\ respectively, are prominent in the IRTF SpeX spectrum, as they are in the ISO SWS spectrum, and have full widths at half-maximum (FWHMs) $\sim$ 0.008~\mum. Weak \ion{H}{1} Humphreys ({\it n}$-$6) series lines from upper levels {\it n} = 18$-$24 are also detected; several of these were also noted by \citet {Hensley20}. Higher lines in the series are below the noise level of the spectrum, although the 25$-$6 line at 3.483~\mum\ may contribute to the spectral structure seen in the 3.4~\mum\ hydrocarbon band near that wavelength. We identify a weak emission line at 3.092~\mum\ as \ion{He}{2} 7$-$6; it is also present in the ISO spectrum published by \citet {Hensley20}, although not identified by them. Spurious, sharp spectral structure at 3.31$-$3.32~\mum\ due to the inability to properly correct for the opaque peak absorption of the Q branch of the methane $\nu_{3}$ band in the spectra of both Cyg OB2-12 and the telluric standard, is left out of Figure  \ref{fig:fig1}. In our subsequent analysis and modeling of the hydrocarbon features we exclude the spectral region 3.284$-$3.324~\mum, which contains these false features and the nearby \ion{H}{1} 9$-$5 absorption line. The spectra in subsequent figures include single data points at each end of this excluded interval, but those two points are not included in the modeling. 

 Recently, \citet[][see their Figure 4]{potapov2021dust} presented evidence for a weak ($\tau \sim$ 0.02) and broad spectral feature centered near 2.9~\mum\ in the ISO optical depth spectrum of Cyg OB2-12, which they identify as water trapped in silicate grains. Our ground-based spectrum of Cyg OB2-12, which extends only down to 2.86~\mum, is not conducive to detecting this feature. 
 
\begin{figure*}[!hbt]
\epsscale{1.1}
\includegraphics[trim={0 0 0 0},clip,width=1\textwidth]{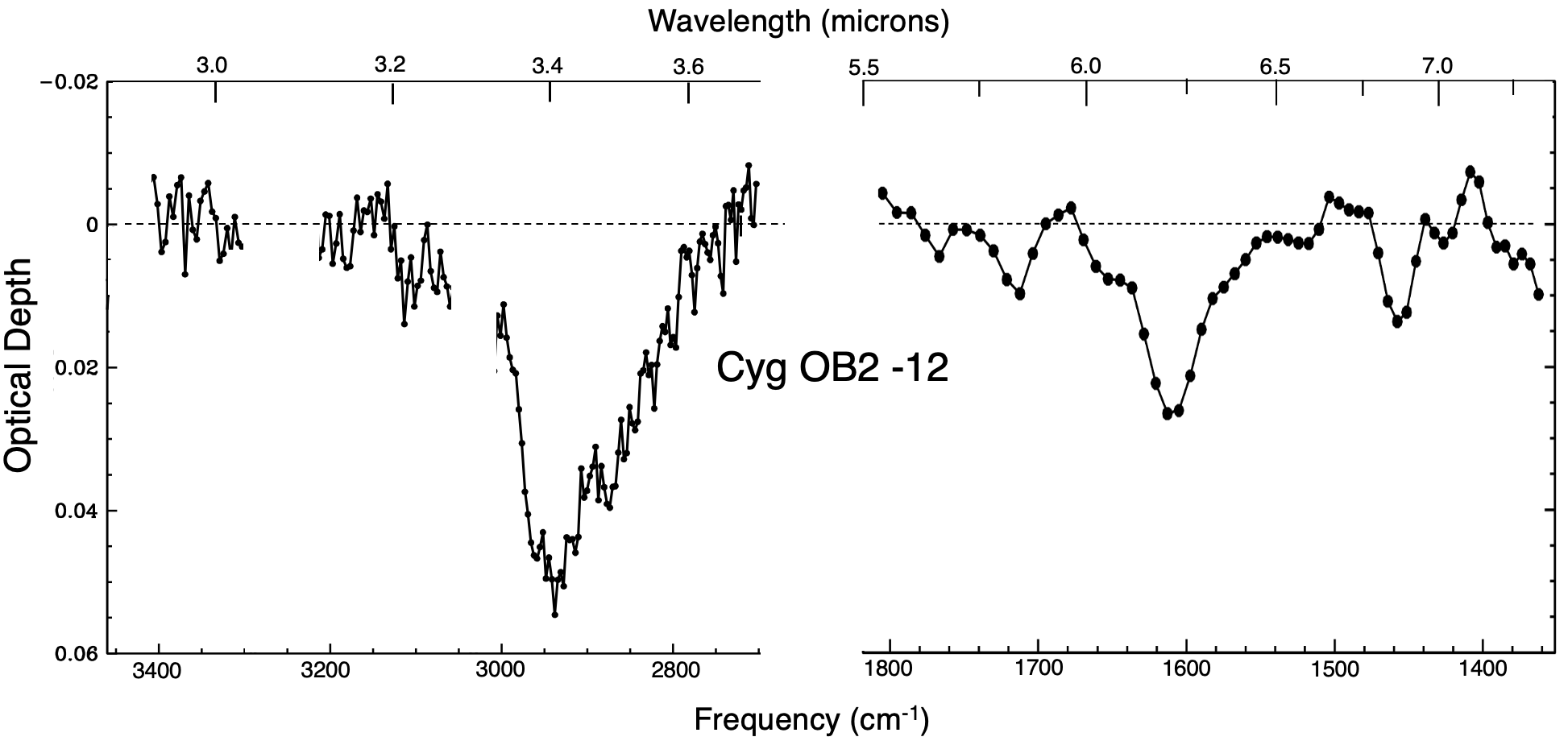}
\caption{The 2.90$-$3.70~\mum\ (NASA IRTF) and 5.50$-$7.34~\mum\ (Spitzer IRS) spectra of Cyg OB2-12 on a common optical depth scale. The resolutions of the two spectra are 0.004 \mum\ and 0.09 \mum, respectively. The gap at 3.025$-$3.10~\mum\ is due to interference by the \ion{H}{1} 10$-$5 absorption and \ion{He}{2} 7$-$6 emission lines. The gap at 3.29$-$3.32~\mum\ is due to interference by the \ion{H}{1} 9$-$5 absorption line at 3.297~\mum\ and nearly total absorption of the spectrum by telluric methane in the 3.31$-$3.32~\mum\ interval.}
\label{fig:fig2}
\end{figure*}

 \subsection{Cyg OB2-12 Mid-infrared}
 \label{sec:Cyg_MIR_obs}
Cyg OB2-12 was observed on 2008 August 6 using the infrared spectrograph (IRS) on the \textit{Spitzer Space Telescope} (PI: David Ardila, AOR 27570176; \cite{ardila2010spitzer}). The spectroscopic data were obtained using the Short-Low (SL) module in all orders (covering 5.2$-$14~\mum) as well as the second order of the Long-Low (LL) module (covering 14$-$21.3~\mum ). In this paper we use a portion of the calibrated 5.2$-$14~\mum\ spectrum derived from these observations by \cite{potapov2021dust}. The resolution of the heavily oversampled spectrum is 0.09~\mum, but here we display the spectrum in bins of 0.03~\mum, which matches the sampling of the mid-IR spectrum of the Quintuplet in \citet{chiar2013structure}.  For a detailed discussion of the calibration and the spectral extraction of the \textit{Spitzer} low-resolution spectrum, see \cite{potapov2021dust}.

 We focus on  
 the 5.5$-$7.34~\mum\ spectral region, which contains the 
 6.2~\mum\ and 6.85~\mum\ hydrocarbon absorption bands, the 5.85 and 6.05~\mum\ bands which we attribute to the C = O bonds, and part of the CH$_3$ symmetric deformation mode at 7.27~\mum. Beyond 7.34~\mum\ the spectrum is unhelpful due to the presence of the strong hydrogen Pf $\alpha$ emission line at 7.46~\mum.  As we did for the shorter wavelength SpeX spectrum, we ratioed the IRS F$_{\lambda}$ spectrum of Cyg OB2-12 to that of a 13,700 K blackbody. However, unlike the ratioed SpeX spectrum, the continuum in this longer wavelength region retains considerable curvature because it contains a large contribution from the stellar wind, which does not have a Rayleigh-Jeans wavelength dependency \citep[see e.g.][]{fogerty2016silicate}. To flatten the continuum, we divided the IRS F$_{\lambda}$ spectrum by a quadratic function tied to the mean values of the curved spectrum near 5.7, 6.6, and 7.2~\mum. None of the narrow intervals centered on these wavelengths correspond to known solid-state absorption bands in diffuse clouds \citep[see e.g.][]{van2011}. 
 
 The resulting spectrum, now with a flat continuum, is shown on the right in Figure \ref{fig:fig2}. In the 6.4$-$6.8~\mum\ interval the spectrum shows some marginally significant broad structure, which could be interpreted as a weak absorption band at 6.4$-$6.6~\mum\ or a weak emission band at 6.6$-$6.8~\mum, depending on the location and value of this central continuum point. Because we do not expect spectral features at 6.4$-$6.8~\mum, we chose as the continuum at 6.6~\mum\ the average value of the spectrum in this rather broad interval. 
 
Three absorption features are evident in the mid-IR spectrum (Figure \ref{fig:fig2}), centered at approximately 5.85, 6.2,  and 6.85~\mum. The 6.2~\mum\ band has a prominent blue shoulder centered at about 6.0~\mum, which we regard as evidence for a fourth absorption band. In addition, \citet{chiar2013structure} treated the 6.2~\mum\ feature toward the Quintuplet as a blend of CC olefinic (6.19~\mum) and aromatic  (6.25~\mum) bands, and we have taken the same approach.  The CH$_3$ symmetric deformation mode at 7.27~\mum\ is not clearly detected. In our analysis we use only the 5.5$-$7.0~\mum\ portion of the mid-IR spectrum of Cyg OB2-12, which we thus consider to be composed of five absorption bands.  The placement of the 6.4$-$6.8~\mum\ continuum level discussed above has a modest effect on the derived strengths of these bands. Although the 6.4$-$6.8~\mum\ interval is shown in the figures, it was not included in the Gaussian fitting due to its ambiguous continuum level.

\newpage

\begin{figure}[!hbt]
\epsscale{1.2}
\includegraphics[clip, trim=0 0 0 0, width=\linewidth]{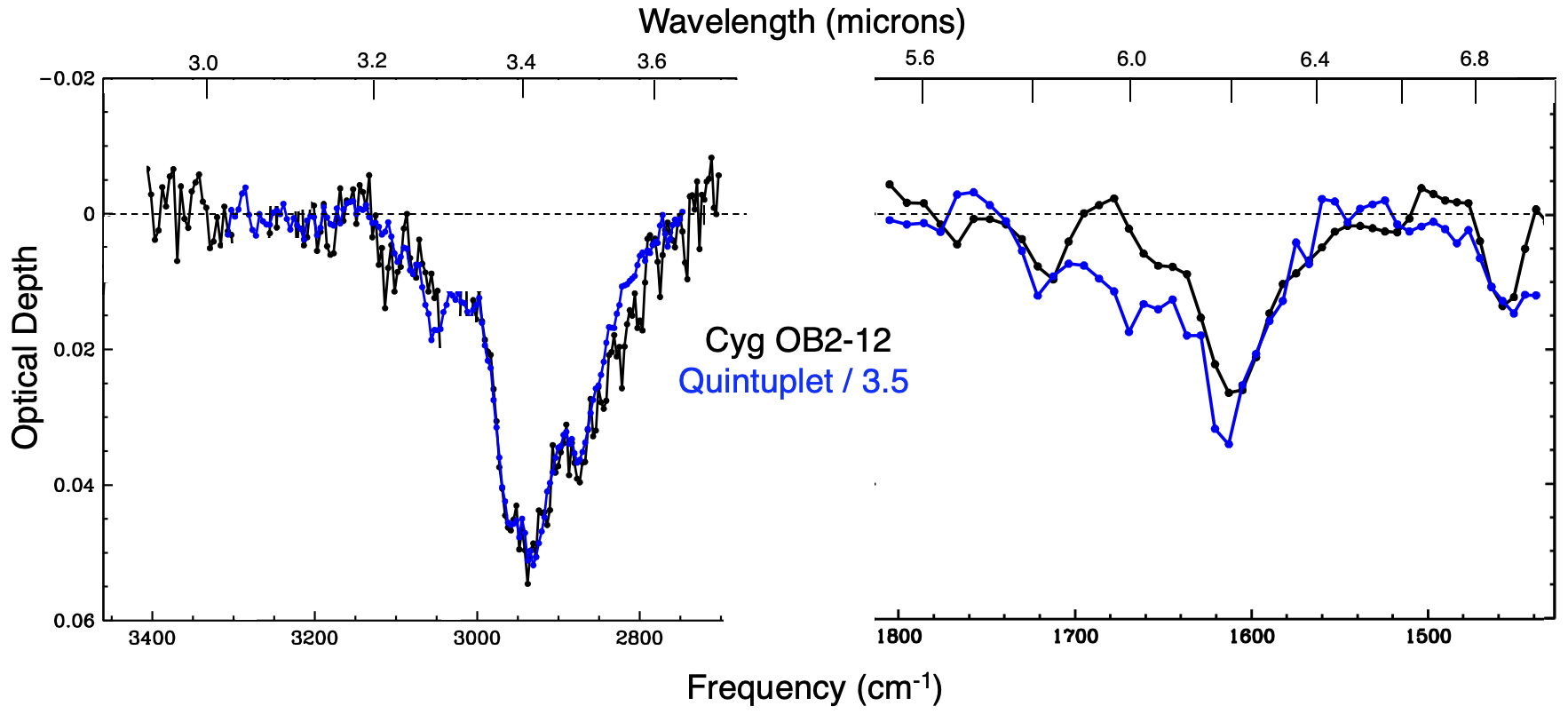}
\caption{Optical depth spectra of Cyg OB2-12 (black) and Quintuplet (blue) sightlines.  Both the near-IR and mid-IR optical depth spectra of the Quintuplet \cite[both from][]{chiar2013structure} are divided by 3.5 so that the strengths of the 3.4 \mum\ match and thus facilitate comparison of relative band intensities. The continuum level of the mid-IR spectrum of the Quintuplet is slightly shifted from that in \cite{chiar2013structure} as discussed in the text.}
\label{fig:fig3}
\end{figure}

\subsection{Comparison of Cyg OB2-12 and Quintuplet Spectra}
\label{sec:Cyg_Q_obs_comp}

Figure \ref{fig:fig3} shows the near-IR and mid-IR optical depth spectra toward Cyg OB2-12 co-plotted with the spectra toward the Quintuplet. The spectra of both wavelength regions for the Quintuplet were published by \citet{chiar2013structure}; however, those authors only published the 3.20-3.64 \mum\ portion of the near-IR spectrum, whereas we show their spectrum from 3.02 \mum\ to 3.64 \mum. Note that the third-order polynomial they used to fit the near-IR continuum effectively removed any contribution by the long wavelength wing of the weak ($\tau$ $\sim$ 0.2) \mum\ water-ice absorption band toward the Quintuplet sources \citep[see Figure 1 of][]{chiar2000composition}.

As discussed by \citet{chiar2013structure}, the Quintuplet mid-IR spectrum contains a contribution 
near 6~\mum\ by the H$_2$O-ice absorption band. They estimated its strength from the 3~\mum\ ice band in \citet{chiar2000composition} and subtracted it from the  
mid-IR spectrum. We have adopted  their resulting mid-IR optical depth spectrum; however, unlike their choice of continuum \citep[described in][]{chiar2000composition}, which seems to imply that there are broad emission features centered near 5.65 \mum\ and 6.50 \mum, we have defined the continuum for the Quintuplet sightline as a linear fit between the average signals in the 5.5$-$5.7~\mum\ and 6.4$-$6.7~\mum\ intervals. Doing so increases the peak optical depths of absorption features in the mid-IR spectrum by $\sim$ 0.015 compared to those in Fig. 2 of \citet{chiar2013structure}. In our estimation, this approach is a more accurate portrayal of the optical depths of the hydrocarbon bands. Our revised spectrum, resampled in bins of 0.03~\mum, is shown here.

The spectra in the 3.2 $-$ 3.6\,\mum\ interval reveal an extraordinarily close match between the profiles of the bands produced by the CH stretching of aromatic (3.3~\mum) and aliphatic (3.4~\mum) hydrocarbons  along both sightlines. While not as close a match, the absorption features in the 5.5$-$7.0\,\mum\ spectra also resemble one another in their peak wavelengths and relative band depths. Spectra on both sightlines show an absorption band attributed to the C=O stretch in carbonyl groups at approximately 5.85~\mum. In addition, both contain absorption bands due to the aromatic C-C stretch at approximately 6.20~\mum, and the CH$_{2}$ scissor (bending) mode of aliphatic hydrocarbons at 6.85~\mum.

The most significant difference between the spectra along the two sightlines is the additional absorption in the 5.85$-$6.15 \mum\ interval toward the GC Quintuplet region. That wavelength interval approximately corresponds to the center of the $\nu_2$ vibration in H$_2$O-ice. As mentioned above and discussed more fully in Section \ref{sec:Q_LOS}, 
the sightline to the Quintuplet passes through molecular clouds associated with three intervening Galactic spiral arms, producing the weak ice absorption feature at 3.0~\mum\ observed by \citet{chiar2000composition}. Using this ice band those authors constrained and removed the ice absorption contribution in the 6 \mum\ region, thereby producing the mid-IR spectrum for the Quintuplet that is shown in Figure \ref{fig:fig3}. Thus, it appears that residual absorption by water-ice should not cause the extra absorption at $\sim$6.0~\mum.  Absorption lines in the $\nu_2$ band of gaseous H$_2$O can also be prominent spectral features in dense clouds. However, the center of that band is at 6.27~\mum\ and, in addition, the few lines in that band detected by \citet{moneti2001cold} toward the Quintuplet are too weak to noticeably affect the low-resolution spectrum presented here. In summary, we are unable to account for the difference in the spectra at $\sim$6.0~\mum.  The excess $\sim$6.0~\mum\ absorption toward the Quintuplet may reflect a genuine difference in the composition of the dust along these two sightlines.

\begin{figure*}[!htb]
\epsscale{1.2}
\includegraphics[trim={0 0 0 0},clip,width=1\textwidth]{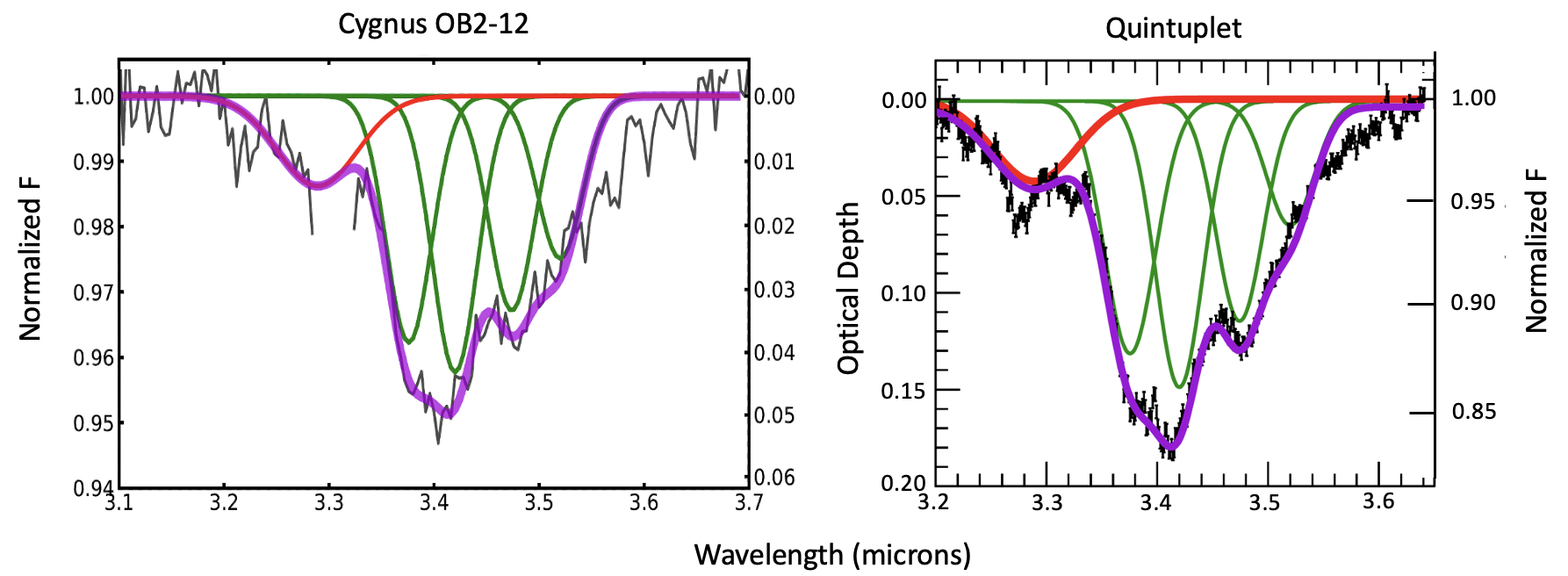}
 \caption{\label{fig:fig4} Normalized flux (F) (left and right y axes) and optical depth ($\tau$) (middle y axes) 3.1$-$3.7 \mum\ spectra of Cyg OB2-12 (left) and Quintuplet (right) (observations shown by black solid lines) with Gaussian fits to the 3.289 \mum\ (aromatic CH) in  red and to  the aliphatic CH stretch bands in green at 3.376 \mum\ (CH$_{3}$ asymmetric), 3.420 \mum\ (CH$_{2}$ asymmetric), 3.474 \mum\ (CH$_{3}$ symmetric), and 3.52 \mum\ (CH$_{2}$ symmetric). The Quintuplet spectrum (a composite of ISO and Spitzer data) and the fit to it are directly from \cite{chiar2013structure} (their Figure 2). The sums of the Gaussian fits are shown by purple lines for both sightlines. Details of the fits are summarized in Table \ref{tab:t1}.} 
\end{figure*}

\begin{figure*}[!htb]
\epsscale{1.2}
\includegraphics[trim={0 0 0 0},clip,width=1\textwidth]{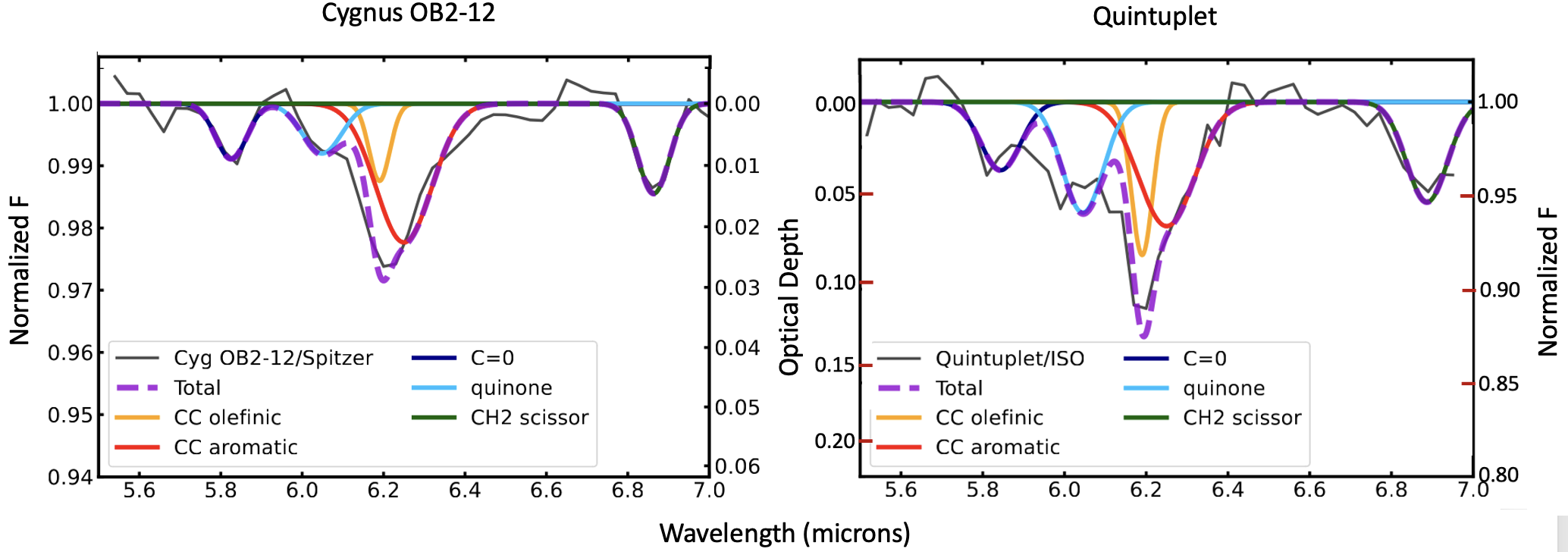}
\caption{Normalized flux (F) (left and right axes) and optical depth ($\tau$) (middle axes) spectra are depicted by thin black lines for Cyg OB2-12 and the Quintuplet sources, left and right panels, respectively, over the 5.5$-$7.0 \mum\ range. Gaussian fits to the 5.85 (C=O; blue), 6.05 (quinone; aqua), 6.19 (olefinic; yellow), 6.25 (aromatic; red), and 6.85 (CH$_{2}$ scissor; green) \mum\ absorption bands are color coded as listed in the legend. The sums of the Gaussians are shown by dashed purple lines.} 
\label{fig:fig5}
\end{figure*}

\begin{table*}
\begin{center}
\scriptsize
\caption{\label{tab:t1} Gaussian Fits to Carbonaceous Absorption Bands toward Cyg OB2-12 and the Quintuplet Cluster\tablenotemark{a,b}}
\begin{tabular}{lccccccc}
\hline\hline
Sightline and Mode & $\lambda$\tablenotemark{c} & FWHM & Strength & FWHM & Amplitude & Integrated Area\tablenotemark{d} & Column Density\tablenotemark{d,e} \\
& \mum\ & \mum\ & $10^{-17}$ cm/group & cm$^{-1}$ & of Gaussian & (IA) cm$^{-1}$ & 10$^{18}$ group/cm$^{2}$ \\
\hline \hline
Cyg OB2-12 CH ({\it sp$^2$}) aromatic\tablenotemark{f} & 3.289 & 0.09 & 0.258 & 87.1 & 0.0137 $\pm$ 0.0025 & {\bf1.27}$\pm${\bf0.23} & {\bf0.49} $\pm$ {\bf0.09} \\
Quintuplet CH ({\it sp$^2$}) aromatic\tablenotemark{f} & 3.289 & 0.09 & 0.258 & 81.8 & 0.042 & {\bf 3.7}\tablenotemark{g} &{\bf 1.43} \\
    \hline
Cyg OB2-12 CH$_3$ ({\it sp$^3$}) asym. & 3.376 & 0.05 & 2.43 & 50.3 & 0.0386 $\pm$ 0.0025 & 2.07 $\pm$ 0.13 & 0.085 $\pm$ 0.006 \\
Cyg OB2-12 CH$_2$ ({\it sp$^3$}) asym. & 3.420 & 0.05 & 1.52 & 45.8 & 0.0433 $\pm$ 0.0024 & 2.12 $\pm$ 0.12 & 0.139 $\pm$ 0.008 \\
Cyg OB2-12 CH$_3$ ({\it sp$^3$}) sym.  & 3.474 & 0.05 & 2.37 & 43.7 & 0.0331 $\pm$ 0.0024 & 1.54 $\pm$ 0.11 & 0.065 $\pm$ 0.005 \\
Cyg OB2-12 CH$_2$ ({\it sp$^3$}) sym.  & 3.520 & 0.05 & 1.48 & 43.0 & 0.0252 $\pm$ 0.0023 & 1.16 $\pm$ 0.11 & 0.078 $\pm$ 0.007 \\
Cyg OB2-12 Total ({\it sp$^3$}) aliphatic& & & & & & {\bf6.89} $\pm$ {\bf0.24} & {\bf 0.18 $\pm$ 0.04}\tablenotemark{h} \\ 
\hline
Quintuplet CH$_3$ ({\it sp$^3$}) asym. & 3.376 & 0.05 & 2.43 & 47.2 & 0.131 & 6.55 & 0.27 \\
Quintuplet CH$_2$ ({\it sp$^3$}) asym. & 3.420 & 0.05 & 1.52 & 42.8 & 0.148 & 6.74 & 0.44 \\
Quintuplet CH$_3$ ({\it sp$^3$}) sym.  & 3.474 & 0.05 & 2.37 & 41.0 & 0.115 & 4.96 & 0.21 \\
Quintuplet CH$_2$ ({\it sp$^3$}) sym.  & 3.520 & 0.05 & 1.48 & 40.4 & 0.064 & 2.76 & 0.19 \\
Quintuplet Total ({\it sp$^3$}) aliphatic  & & & & & & {\bf 21.01} & {\bf 0.56} $\pm$ {\bf 0.08}\tablenotemark{h} \\ 
\hline
Cyg OB2-12 C=O & 5.826 & 0.090 & 1.25 & 26.5 & 0.0089 $\pm$ 0.0010 & 0.25 $\pm$ 0.03 & 0.020 $\pm$ 0.002 \\
Quintuplet C=O & 5.844 & 0.113 & 1.25 & 33.1 & 0.0372 $\pm$ 0.0032 & 1.31 $\pm$ 0.11 & 0.105 \\
\hline
Cyg OB2-12 quinone-like\tablenotemark{i} & 6.049 & 0.113 & 1.25 & 30.9 & 0.0081 $\pm$ 0.0008 & 0.27 $\pm$ 0.03 & 0.022 $\pm$ 0.002 \\
Quintuplet quinone-like\tablenotemark{i} & 6.045 & 0.117 & 1.25 & 32.0 & 0.0610 $\pm$ 0.0032 & 2.08 $\pm$ 0.11 & 0.17 $\pm$ 0.01 \\
\hline
Cyg OB2-12 CC ({\it sp$^2$}) olefinic & 6.19 & 0.060 & 0.0275 & 16.0 & 0.0125 $\pm$ 0.0014 & 0.213 $\pm$ 0.024 & 0.77 $\pm$ 0.08 \\ 
Quintuplet CC ({\it sp$^2$}) olefinic & 6.19 & 0.060 & 0.0275 & 15.7 & 0.0853 $\pm$ 0.0056 & 1.42 $\pm$ 0.01 & 5.2 $\pm$ 0.4 \\ 
\hline
Cyg OB2-12 CC ({\it sp$^2$}) aromatic & 6.25 & 0.160 & 0.0275 & 41 & 0.0226 $\pm$ 0.0010 & {\bf 0.99} $\pm$ {\bf0.04} & {\bf 3.60} $\pm$ {\bf0.15} \\
Cyg OB2-12 Total CC ({\it sp$^2$})\tablenotemark{j} & & & & & & 1.20 $\pm$ 0.05 & 4.4 $\pm$ 0.2 \\
Quintuplet CC ({\it sp$^2$}) aromatic & 6.25 & 0.160 & 0.0275 & 41 & 0.0687 $\pm$ 0.0037 & {\bf3.00 $\pm$ 0.16} & {\bf10.9} $\pm$ {\bf 0.6} \\ 
Quintuplet Total CC ({\it sp$^2$})\tablenotemark{j} & & & & & &  4.42 $\pm$ 0.19 & 16.1 $\pm$ 0.7 \\
\hline
Cyg OB2-12 CH$_2$ scissor aliphatic & 6.864 & 0.090 & 0.239 & 19.1 & 0.0146 $\pm$ 0.0010 & 0.30 $\pm$ 0.02 & 0.125 $\pm$  0.008 \\
Quintuplet CH$_2$ scissor  aliphatic & 6.887 & 0.116 & 0.239 & 24.5 & 0.0551 $\pm$ 0.0033 & 1.44 $\pm$ 0.09 & 0.60 $\pm$ 0.04 \\
\hline
Cyg OB2-12 silicate\tablenotemark{k} & 9.7 & 2.15 & 12 & 229 & 0.52 $\pm$ 0.01 & {\bf 119} $\pm$ {\bf 2} & {\bf 0.99} $\pm$ {\bf 0.02} \\
Quintuplet silicate\tablenotemark{k} & 9.7 & 2.15 & 12 & 220 & 2.38 & {\bf 524} & {\bf 4.4} \\
\hline
\end{tabular}
\end{center}
\scriptsize
$^{a}$ General note: ``IA'' refers to the integrated area of the Gaussian ($\approx$ 1.067 $\times$ amplitude $\times$ FWHM in cm$^{-1}$); ``Strength'' refers to the integrated absorption coefficient per absorbing group, in units of $10^{-17}$~cm per group, based on laboratory measurements or values adopted from \citet{chiar2013structure}; ``sp$^2$'' and ``sp$^3$'' indicate the hybridization state of carbon atoms, associated with double and single bonding, respectively; ``asym.'' and ``sym.'' denote asymmetric and symmetric vibrational modes; FWHM values are listed both in microns (\mum) and in wavenumbers (cm$^{-1}$), where the latter refer to the widths used in the optical depth profile fits.\\
$^{b}$ Fits to Quintuplet data and values in this table are from \citet[][uncertainties in their values were not published]{chiar2013structure}; all others this paper. Additional uncertainties in the values for the mid-IR bands due to placement of the continuum. For Cyg OB2-12 we estimate these to be C=O: 25\%\, quinone: 25\%, CC olefinic: 15\%, CC aromatic: 10\%, CH$_{2}$ scissor: 15\%. For the Quintuplet we estimate them to be C=O: 20\%, quinone: 15\%, CC olefinic: 10\%, CC aromatic: 10\%, CH$_{2}$ scissor: 20\%. \\
$^{c}$ Central wavelength of Gaussian. \\
$^{d}$ Boldface values in this column indicate that these entries were used to calculate values in Tables \ref{tab: Ratios_Cyg_Q} and \ref{tab:dust mass}.\\
$^{e}$ Column densities were calculated by dividing the integrated area (IA, in cm$^{-1}$) by the band strength (in cm per group), using $N = \mathrm{IA} / A$, where $A$ is the integrated absorption coefficient (column "Strength") reported in units of $10^{-17}$~cm per group. For example, for the 3.3~\mum\ aromatic CH mode in Cyg OB2-12, $N = 1.27~\mathrm{cm}^{-1} / (0.258 \times 10^{-17}~\mathrm{cm}) = 4.92 \times 10^{17}~\mathrm{groups/cm}^2$, or 0.49  in units of $10^{18}$~groups/cm$^2$. Where multiple CH$_2$ or CH$_3$ components were present, total column densities were computed as the sum of the means of the symmetric and asymmetric values (see note $^g$). Uncertainties were propagated assuming independent errors in IA and strength. \\
$^{f}$ Elsewhere in this paper this absorption band is designated the 3.3 \mum\ band. For the Quintuplet the Gaussian fit (Figure \ref{fig:fig4}, right panel) is to the broad aromatic mode only (we do not fit the superposed narrow ($\sim$25 cm$^{-1}$) component centered at 3.28~\mum). \\
$^{g}$ Corrected typo in Table \ref{tab:t1} of \citet{chiar2013structure} which is listed there as ten times smaller.\\
$^{h}$ Sum of the  mean of the CH$_3$ symmetric and asymmetric column density values and the mean of the CH$_2$ symmetric and asymmetric values (example using Cyg OB2-12: Total (sp$^3$) aliphatic = [(0.085 + 0.065)/2] +[(0.139+0.078)/2]= 0.18). \\
$^{i}$ For discussion of identification see \S~\ref{sec:Disc}.\\
$^{j}$ For each sightline, Total CC value is the sum of the 6.19~\mum\ and 6.25~\mum\ integrated areas (IAs). Bolded IA and column density values were used to calculate the mass and volume ratios in Table~\ref{tab:dust mass}. 
 \\
$^{k}$ All silicate values refer only to the 9.7 \mum\ feature, and all except for column density and IA are taken from \cite{Hensley20}, \cite{whittet1997infrared}, \citet{chiar2000composition}, and \cite{tielens1987composition}. In the amplitude column, the entries are optical depths; for the Quintuplet we use the value for the GC3 sightline in \citet{chiar2000composition}. The values of IA are not derived from Gaussian fits, as the silicate profiles do not resemble Gaussians. Instead, they are approximated as the product of the FWHM and the peak optical depth. This box-shaped estimate, while crude, is sufficient for the comparative purposes of this paper.\\
\end{table*} 

\newpage

\subsection{Gaussian Fits}
\label{sec:Gauss_fits}
 In Figure \ref{fig:fig4}, the near-IR spectra of the Cyg OB2-12 and Quintuplet sightlines are shown together with our single- and multiple-Gaussian fits to the absorption bands. The right panel of Figure \ref{fig:fig4} is taken directly from \citet[][their Figure 2]{chiar2013structure}. For Cyg OB2-12 we have elected to use the same central wavelengths for the five Gaussians as they did, and vary the FWHMs to produce the best match. Figure \ref{fig:fig5} provides a similar comparison of the mid-IR spectra and Gaussian fits for both sightlines. Here we also have used five Gaussians to fit the mid-IR spectra (see the discussion in Section \ref{sec:Cyg_MIR_obs}).
 
 Table \ref{tab:t1} lists the parameters of Gaussian fits: for the 3.3~\mum\ absorption (single Gaussian), the 3.4~\mum\ absorption band (four Gaussians), the 5.8~\mum\ (single) and 6.05~\mum\ (single) absorptions, the 6.2~\mum\ absorption band  \citep[we follow][and fit this feature with two Gaussians centered at 6.19 \mum\ and 6.25 \mum]{chiar2013structure}, and the 6.85~\mum\ absorption (single). The table also includes parameters of the 9.7~\mum\ silicate band, which was not fitted, but is used in our later analysis (Section~\ref{sec:dust variations}) to compare the relative abundances of carbonaceous and silicate dust components along each sightline.

 As can be seen in these figures, good matches to the observed spectra have been obtained. However, the fits do not reproduce the long wavelength wing of the 3.4 \mum\ band at 3.55$-$3.60 \mum, present on both sightlines. This additional absorption might be due to a CH group on a tertiary carbon in a diamond-like structure \citep{allamandola1993diamonds}. We also note that despite the improved quality of this spectrum of Cyg OB2-12 compared to previously published spectra, the detailed profile of its 3.3~\mum\ feature remains fairly uncertain due to its weakness and consequent low signal-to-noise ratio, as well as the interference by the \ion{H}{1}, {\it n} = 9$-$5 line. (We note that this band, observed toward the GC object 2MASS J17470898–2829561, was fitted by a composite of four Gaussians by \cite{bernstein2024analysis}).

An absorption feature at 6.2 \mum\ in the ISO spectrum of Cyg OB2-12 has been identified as water trapped in silicates by \citet[][their Figure 4]{potapov2021dust}. This feature appears to be the same one as in the Spitzer spectrum of Cyg OB2-12 presented here. Our Gaussian fit to it in Figure \ref{fig:fig5} using the wavelengths of the olefinic and aromatic CC modes at 6.19 and 6.25 \mum, identified by \citet{chiar2013structure} in the spectrum of the Quintuplet, fits the Spitzer spectrum well.  
Together with the similar relative strengths of the 6.2 \mum\ absorption and the other hydrocarbon absorptions in our Cyg OB2-12 and Quintuplet spectra, this suggests that 
the majority of the observed 6.2 \mum\ absorption arises from hydrocarbons. Nevertheless, a contribution from trapped water cannot be excluded and may affect the measured band strength toward Cyg OB2-12.

\section{Analysis and Results}
\label{sec:Analysis}

The infrared hydrocarbon bands observed on the Cyg OB2-12 and the Quintuplet sightlines arise from specific molecular structures, including methyl (CH$_3$), methylene (CH$_2$), olefinic (C=C), and aromatic (ring-based) carbon groups, as well as oxygen-bearing species such as carbonyl (C=O) and hydroxyl (-OH). These functional groups are summarized in Table~\ref{tab:t2}. 

In Section \ref{sec:sightlines and chemistry} we provide detailed descriptions and discussions of the complex sightlines to Cyg OB2-12 and the Quintuplet. In the following Section \ref{locations} we present evidence constraining the physical environments (dense or diffuse clouds) in which the carriers of the aliphatic and aromatic bands are found on these and other sightlines.  In Section \ref{sec:dust variations}, we compare the abundances of the aliphatic, aromatic, and olefinic groups responsible for the observed absorption bands to the $K$-band extinction, the silicate feature, and each other; these comparisons provide some further constraints on the locations of the aromatic and aliphatic hydrocarbons on the Quintuplet sightline. Section \ref{sec:dust mass} contains estimates of the masses and volumes per hydrogen atom and per silicate compound of each of these molecular groups.

\begin{deluxetable*}{lp{4.8in}}
\tablecaption{Overview of Functional Group Classes Relevant to Infrared Absorption Features \label{tab:t2}}
\tablehead{
\colhead{\textbf{Class}} & \colhead{\textbf{Structural Characteristics}}
}
\startdata
\textbf{Aliphatic (sp$^3$)} & Carbon atoms connected by single bonds to four other atoms in a tetrahedral arrangement, forming flexible, chain-like structures. The chains often include methyl (–CH$_3$) and methylene (–CH$_2$–) groups. \\
\textbf{Aromatic (sp$^2$)} & Carbon atoms arranged in stable ring structures where the electrons are evenly spread across the rings. These flat, hexagonal rings often occur in large clusters in interstellar carbon dust and are known as polycyclic aromatic hydrocarbons (PAHs). Each carbon is bonded to three other atoms in a planar arrangement. \\
\textbf{Olefinic (sp$^2$)} & Unsaturated hydrocarbon compounds with the general formula C$_n$H$_{2n}$ connected by at least one carbon double bond (C=C). These structures may act as links between aliphatic or aromatic units or appear independently in the carbon network. The carbon atoms involved in double bonds are also bonded to three atoms in a flat configuration. \\
\textbf{Carbonyl} & Functional groups containing carbon atoms double-bonded to oxygen atoms (C=O). Found in several oxygen-containing compounds such as: (i) aldehydes(C=O at the ends of carbon chains), (ii) ketones (C=O within carbon chains), and (iii) quinone-like structures (C=O bonded to an aromatic ring where the double bonds are rearranged to maintain conjugation). In some structures, such as unsaturated conjugated ketones, carbonyl groups are linked to carbon-carbon double bonds (C=C), affecting the infrared absorption properties. These groups absorb in the 5–6 \mum\ region. \\
\enddata

\tablecomments{
This table outlines the major types of carbon-based functional groups responsible for infrared absorption features observed in interstellar dust. 
The terms sp$^3$ and sp$^2$ refer to different bonding geometries: sp$^3$ carbons form four single bonds in a tetrahedral shape, while sp$^2$ carbons form three bonds in a flat (planar) structure. Some entries, such as the carbonyl group, refer to specific functional groups, while others describe broader structural classes of hydrocarbon bonding.
}
\end{deluxetable*}

\subsection{Sightlines}
\label{sec:sightlines and chemistry}

Overall, the close similarity of the Cyg OB2-12 and Quintuplet near- and mid-IR spectra presented here, corresponding to such different sightlines, is striking and is consistent with earlier suggestions that the carriers of the carbonaceous spectral features primarily reside in the  diffuse ISM. However, it is not clearly established for many of these absorption bands that they arise {\it only} in the diffuse ISM. Moreover, there are profound differences in the diffuse environments, not only between the two sightlines, but within each sightline, which make the close agreement even more remarkable. The two sightlines are described and discussed in the following two sections.

\subsubsection{Cyg OB2-12}
\label{sec:Cyg_LOS}
 Thorough discussions of the extinction to Cygnus OB2-12 have been presented by \citet{wright2015massive} and \citet{maryeva2016nature}. \citet{wright2015massive} found that the Cygnus OB2 association contains at least 169 OB stars, which suffer a wide range of visual extinctions, with a mean visual extinction of 5.4~mag. This is roughly half of the total extinction toward Cyg OB2-12. \citet{maryeva2016nature} found that the extinction to association members close in the sky to Cyg OB2-12 increases with decreasing angular separation from Cyg OB2-12, by a total of $\sim~2.5$ mag (see their Fig, 7a), and that approximately 1 magnitude of this increase appears to be associated with the immediate environment of Cyg OB2-12, possibly arising from a circumstellar shell of diffuse dusty gas swept up by the strong stellar wind of this hypergiant \citep{Leitherer1984, Morford2016}. While a modest excess of extinction is present in the immediate vicinity of the star, current infrared observations reveal no emission features indicative of a dense or dusty shell. Taken together, these results suggest that there are at least two, and possibly three, distinct regions of extinction towards Cyg OB2-12, with $\sim~5$ mag located within the association. The remaining $\sim~5$ mag on the Cyg OB2 sightline presumably lie in diffuse gas external to the association. The intense ultraviolet (UV) radiation field within some or all of the cluster may inhibit, to varying degrees, the formation of the hydrocarbons responsible for the absorption features being studies in this paper, and may do so while not being hostile to the silicate particles that are responsible for the extinction.

\subsubsection{Quintuplet}
\label{sec:Q_LOS}

The sightline to the Quintuplet sources is also complex, but in ways that differ from the sightline to Cyg OB2-12. First, the Quintuplet sightline passes through dense gas in three intervening spiral arms \citep{Nogueras_Lara_2021}. \citet{whittet1997infrared} estimated that one-third ($\sim$10 mag) of the visual extinction toward objects close to the central supermassive black hole (Sgr A$^{*}$), a region that includes the nearby Quintuplet stars, arises in dense molecular gas in those arms. Spectra of H$_{3}^{+}$ fundamental band lines and the strong lines of the intrinsically weak CO 2$-$0 band originating from the few lowest rotational states \citep{oka2005cmz,oka2019cmz} yield a similar estimate. Both of these species produce narrow absorptions at the radial velocities of the intervening arms. Extinction toward the Quintuplet star GCS3-2 by dense gas in the spiral arms can be estimated from the sum of the CO column densities in the arms \citep[2.15 $\times$ 10$^{18}$$ \, \text{cm}^{-2}$,][]{oka2019cmz}. Canonical values of [CO]/[H$_{2}$] in dense clouds \citep[e.g.,][]{lee1996} yield dense cloud extinction consistent with the \citet{whittet1997infrared} estimate. In view of this rough agreement, in the following analysis we adopt a diffuse cloud visual extinction of 20 mag toward the Quintuplet.

Second, the sightline to the Quintuplet also passes through the front half of the Central Molecular Zone (CMZ) of the Galaxy \citep{morris1996araa}, which is a disk-like region of $\sim$150 pc radius and thickness several tens of pc centered on Sgr A$^{*}$. In terms of volume, the dominant environment within the CMZ is diffuse ($<n>$ $\sim$ 50 cm$^{-3}$) gas at a mean kinetic temperature of $\sim$ 200 K \citep{oka2019cmz}. The density of hydrogen in the diffuse gas of the CMZ and the filling factor of that gas imply a visual extinction toward the Quintuplet produced by dust in the CMZ of $\sim$ 8 mag, assuming the standard gas-to-dust ratio \citep{savage1977}; this is nearly half of the total extinction by diffuse gas on the Quintuplet sightline. Analysis of spectral lines of H$_3$$^+$ in the CMZ implies that the region is subject to a mean cosmic ray ionization rate two orders of magnitude higher than its average rate in diffuse clouds outside of the GC \citep{oka2019cmz}. The mean UV field in the CMZ is unknown; locally it must be strongly dependent on proximity to clusters of hot stars such as the Quintuplet Cluster. While it is likely that silicate dust survives in this environment, it is unclear if that is the case for hydrocarbon dust (see Section \ref{sec:dust variations}).

\begin{figure}[!htb]
\label{fig:fig6}
\epsscale{0.8}
\includegraphics[trim={0 0 0 0},clip,width=0.47\textwidth]{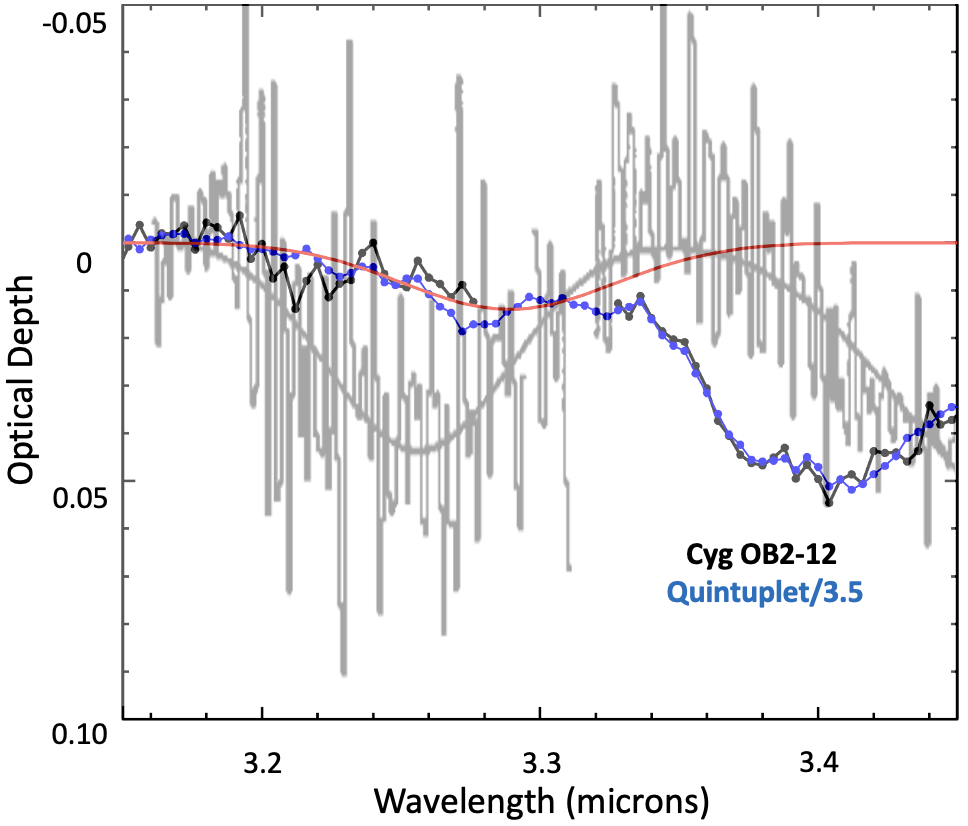}
\caption{\label{fig:fig6}
Comparison of spectral profiles of the 3.3 \mum\ absorption bands towards Cyg OB2-12 (black) and the GC Quintuplet (blue, scaled down by a factor of 3.5; data from Figure \ref{fig:fig3} of this paper), and the 3.25 \mum\ band in the deeply embedded YSO Mon R2 IRS3 \citep[][shown with its Gaussian fit, both in gray]{sellgren95}. The Gaussian used to fit the 3.3 \mum\ bands in the spectra of Cyg OB2-12 and the Quintuplet is shown in red. }
\end{figure}

\subsection{Locations of Carriers of Aliphatic and Aromatic Bands}
\label{locations}

As described earlier, the 3.4~\mum\ aliphatic absorption band has been detected along several Galactic sightlines through the diffuse interstellar medium, but it has not been observed in dense molecular clouds.  Theoretical and laboratory experiments have made a case for why this band is not present in dense clouds: the formation of water-ice mantles suppresses the reformation of CH$_2$ and CH$_3$ groups after the original aliphatic material from the diffuse ISM has been destroyed by energetic processes  \citep[e.g.,][]{mennella2001uv, mennella2008activation}. 

Observationally, the non-detection of the aliphatic 3.4 \mum\ band in dense clouds might be due in part to its location on the steeply rising long wavelength wing of the strong 3~\mum\ water-ice absorption band. This wavelength region often contains  additional absorption features from contaminants within the ice mantles \citep[e.g.,][]{brooke1996study, schutte1996,dartois1999,mcclure2023}, further complicating the detection of weaker hydrocarbon features.  Nevertheless, it can be argued that such observational limitations do not account for the absence of the 3.4~\mum\ band. The extinctions toward embedded sources in dense clouds are typically much higher than those toward diffuse cloud sightlines such as Cyg OB2-12, where the 3.4~\mum\ band is easily detected at an optical depth of $\sim$0.05 (Table \ref{tab:t1} and Figure \ref{fig:fig1}).
Based on optical depths of the 9.7~\mum\ silicate band, the extinctions toward sources in dense molecular cloud cores, e.g. those listed in \citet{willner82} and \citet{sellgren94} are up to an order of magnitude greater  than that toward Cyg OB2-12. If the 3.4~\mum\ band scaled proportionally with the silicate optical depth and/or with optical extinction, one would expect  optical depths of several tenths in dense clouds, which would be readily detectable, especially given the band's characteristic sharp onset near 3.36 \mum\ (e.g., see Figure \ref{fig:fig6}).
 Yet this band has not been detected. Thus, it is generally accepted that the dust responsible for the aliphatic hydrocarbon bands resides exclusively in the diffuse interstellar medium.

Although the 3.3~\mum\ aromatic bands detected toward Cyg OB2-12 and the Quintuplet are considerably weaker than the 3.4~\mum\ aliphatic bands along those sightlines, a similar extinction-based argument also suggests their absence in dense molecular clouds. If the 3.3~\mum\ carrier were as abundant in dense molecular clouds as toward Cyg OB2-12, its strength would scale proportionally with the higher extinctions typically observed along embedded sightlines, implying an expected  optical depth of $\sim$ 0.1$-$0.2. However, spectra of sources embedded in dense clouds (see references above) show little or no evidence for such a band. In a few cases where a possible weak 3.3 \mum\ feature is present, its optical depth is far less than predicted by scaling with extinction.

As pointed out in the Introduction, a weak band centered near 3.25 \mum\ has been detected at low signal-to-noise ratios toward several YSOs embedded in dense molecular clouds \citep[e.g.][]{sellgren95, brooke1996study}. This absorption feature overlaps the aromatic 3.3 \mum\ band observed toward Cyg OB2-12 and the Quintuplet.  The 3.25 \mum\ band has also been attributed by the above authors to CH aromatic stretching vibrations. However, in agreement with \citet{chiar2013structure}, our analysis reveals significant differences between it and the 3.3 \mum\ absorption band detected in the diffuse ISM. The five embedded sources toward which this band has been detected have a mean central wavelength of 3.252~$\pm$~0.010~\mum\ and mean full width at half maximum (FWHMs) of 0.070~$\pm$~0.010~\mum. In contrast, the single-Gaussian fit to the 3.3~\mum\ absorption observed toward the Quintuplet and Cyg OB2-12 a central wavelength of 3.289~\mum\ and a FWHM of 0.09~\mum. The differences between the profiles of the two bands are illustrated in Figure \ref{fig:fig6}.
 Thus, although both dense and diffuse ISM environments exhibit CH aromatic stretching absorptions near 3.3~\mum, the bands in the two environments likely arise from chemically distinct carriers. The 3.3~\mum\ band observed toward Cyg OB2-12 and the Quintuplet appears to be uniquely associated with the diffuse ISM.
 
Future observations with the JWST, which can obtain sensitive spectra of numerous background stars behind and around dense molecular clouds, will provide a critical test of whether hydrocarbon features in the diffuse ISM survive incorporation into denser environments. By targeting faint field stars located behind a dense cloud but viewed through its edge, JWST will be able to probe the transition zone where diffuse material is being accreted into molecular clouds. In this outer region — prior to the onset of ice mantle formation at A$_V$~$\sim$~4 — the destructive processing predicted to erase the aliphatic and aromatic bands \citep[e.g.,][]{mennella2001uv} may not yet have occurred.

\begin{table*}[!hbt]
\begin{center}
\normalsize
\caption{Aromatic and Aliphatic Band Integrated Areas (IAs) and Ratios\tablenotemark{a,b}}
\label{tab: Ratios_Cyg_Q}
\begin{tabular}{l|ccc|c|c}
\toprule
\textbf{Source} & IA(3.3~\mum) & IA(3.4~\mum) & IA(6.25~\mum) & IA(3.3)/IA(3.4) & IA(6.25)/IA(3.4) \\
               & (cm$^{-1}$) & (cm$^{-1}$) & (cm$^{-1}$) & & \\
\midrule
Cyg OB2-12 & 1.27 $\pm$ 0.23 & 6.89 $\pm$ 0.24 & 0.99 $\pm$ 0.04 & 0.184 $\pm$ 0.034 & 0.144 $\pm$ 0.008 \\
Quintuplet\tablenotemark{c} & 3.7 & 21.01 & 3.00 $\pm$ 0.16 & 0.176 & 0.143 $\pm$ 0.008 \\
\bottomrule
\end{tabular}
\end{center}
$^{a}$ Values of IAs from Table~\ref{tab:t1}.\\
$^{b}$ We do not use the 6.05 \mum\ and 6.85 \mum\ bands in this table, as they are different vibrational modes of structural units that are already accounted for; e.g., we use the 6.25 \mum\ band CC aromatic stretch for the aromatic component and the CH$_{2}$ and CH$_{3}$ stretching modes for the aliphatic modes (see footnote g in Table~\ref{tab:t1}).   \\
$^{c}$ Values without uncertainties from \citet{chiar2013structure}; uncertainties were not published. \\

\end{table*}

\subsection{Relative Abundances of Carbon Dust Components and Silicates}
\label{sec:dust variations}

In this section, we compare the ratios of the  aliphatic and aromatic band strengths along the two sightlines. Table~\ref{tab: Ratios_Cyg_Q}  shows these comparisons using the values of the IAs for the aromatic and aliphatic bands in Table~\ref{tab:t1}. As we are comparing ratios between two different sightlines, uncertainties in the conversion from integrated areas to column densities do not affect our conclusions.\footnote{We compare integrated areas (IAs) rather than column densities because IAs are directly measured from the spectra and do not depend on uncertain band strength values. This avoids introducing additional uncertainties associated with converting to column densities. Because we are comparing the same bands across two sightlines, using IA ratios provides a robust basis for interpretation.}

As discussed in Section~\ref{locations}, we present strong evidence that carriers of both the aliphatic and aromatic bands are only present in the diffuse ISM. However, the survivability of hydrocarbon dust in the CMZ portion of the Quintuplet sightline remains uncertain. Laboratory experiments have shown that ion bombardment removes hydrogen atoms from HAC material  \citep{mennella2003effects}. In such environments, the elevated cosmic ray flux may shift the balance between hydrogen loss (from UV and cosmic ray processing) and rehydrogenation via collisions with atomic hydrogen, resulting in a lower H/C ratio in carbonaceous dust in the CMZ. It is also uncertain to what extent  hydrocarbon dust survives within the Cyg OB2 association given the high UV field associated with this environment. 

Following \citet[][their eqn.~14]{mennella2003effects} and \citet[][their eqn.~3]{chiar2013structure}
the balance between hydrogen loss (primarily due to UV photodissociation) and hydrogen addition through collisions with atomic hydrogen is given by
\begin{equation}
\theta = \left( 1 + \Gamma \right)^{-1} \theta_{\mathrm{sat}},
\end{equation}
\noindent
with $\Gamma = \sigma_{\mathrm{UV}}N_{\mathrm{UV}}/\sigma_\mathrm{H}v_\mathrm{H}n_\mathrm{H}$, and
where $\theta$ is the H/C ratio, $\theta_{\mathrm{sat}}$ is the saturation value\footnote{Derived from laboratory studies of HAC analogs, which show that fully hydrogenated carbonaceous solids typically contain one hydrogen atom per two carbon atoms (e.g., \citealt{DuleyWilliams1981}; \citealt{jones2012a}). This value is commonly adopted as an upper limit for hydrogen coverage on interstellar HAC-like grains.} (taken here to be 0.5), $\sigma_{\mathrm{UV}}$ and $\sigma_{\mathrm{H}}$ are the cross sections for UV photo-dissociation and H-atom attachment, respectively, $N_{\mathrm{UV}}$ is the far-UV photon flux, $v_{\mathrm{H}}$ is the thermal velocity of hydrogen atoms, and $n_{\mathrm{H}}$ is the hydrogen nucleus number density.

Assuming representative values for diffuse cloud conditions,\footnote{Specifically, $\sigma_{\mathrm{UV}} \sim 10^{-19}~\mathrm{cm}^2$, $\sigma_{\mathrm{H}} \sim 1.5 \times 10^{-18}~\mathrm{cm}^2$, $v_{\mathrm{H}} \sim 1.5 \times 10^5~\mathrm{cm\,s^{-1}}$, and $N_{\mathrm{UV}} \sim 10^8\, G_0~\mathrm{photons\,cm^{-2}\,s^{-1}}$, following the approximations in \citet{chiar2013structure}.}
the above expression simplifies to:
\begin{equation}
\theta \simeq \frac{0.5}{1 + 30\, G_0 / n_H},
\end{equation}
\noindent where $G_0$ is the strength of the UV radiation field normalized to the Habing field, $10^8$ photons cm$^{-2}$ s$^{-1}$ \citep{Habing1968}, and n$_{\mathrm{H}}$ is the density of hydrogen nuclei in cm$^{-3}$.
For typical diffuse interstellar conditions, $G_0 / n_{\mathrm{H}} \simeq 10^{-2}$ \citep{tielens2005physics}, this expression yields $\theta \simeq 0.38$. In dense clouds, UV photons are produced by cosmic ray excitation of H$_2$ followed by radiative decay in the Lyman-Werner bands \citep{prasadtarafdar1983}, resulting in $G_0 \simeq\ 5 \times10^{-4}$. Rehydrogenation is inhibited by ice mantles and the timescale for H-loss is $\sim$2 Myr, if we also take direct cosmic ray interaction with HAC into account \citep{mennella2003effects}. This is considerably shorter than the typical lifetime of molecular clouds, $\sim 20$ Myr \citep{chevance2023}. We note that in OB associations or other photodissociation regions (PDRs), where $G_0 / n_{\mathrm{H}} \simeq 1$ \citep{tielens2005physics}, $\theta$ falls to approximately 0.016, 
demonstrating the potential strong depletion of hydrogen from HAC in intense UV environments such as the CMZ (c.f., \S~\ref{sec:Q_LOS}).

This behavior indicates that normal diffuse clouds favor the formation of H-rich amorphous carbon (HAC), whereas the harsher conditions of the CMZ, characterized by both strong UV radiation and elevated cosmic ray flux, drive grains toward the H-poor end member of the amorphous carbon spectrum, tetrahedral amorphous carbon (ta-C:H), composed largely of sp$^3$-bonded carbon with very low hydrogen content.
Because HAC and ta-C:H represent structural end members, their spectral signatures—particularly the 3.4 \mum\ band—are relatively insensitive to moderate variations in physical conditions. This is consistent with the observed near-invariance of the 3.4 \mum\ absorption profile across different diffuse ISM sightlines. While local conditions probably influence the hydrogen content of amorphous hydrocarbon mantles -- responsible for the 3.4 \mum\ feature -- the carbon backbone in the HAC/ta-C:H mantle will not be affected. We also expect that the underlying aromatic carbon cores are chemically more resilient and likely retain characteristics inherited from their stellar origin.

The 3.3 and 6.2~\mum\ features are attributed to aromatic carbon stardust. As can be seen in the rightmost two columns of Table \ref{tab:t1} the ratios of integrated areas of aromatic to aliphatic bands, IA(3.3)/IA(3.4) and IA(6.25)/IA(3.4), are virtually the same on the two sightlines. We focus on these two aromatic bands because they are strong, well-isolated, and consistently detected, unlike the 6.05 and 6.85~\mum\ features, which are broader and more susceptible to blending and continuum placement uncertainties.

As already noted in Section \ref{sec:Cyg_Q_obs_comp}, the profiles of these bands are essentially identical on the Cyg OB2-12 and Quintuplet sightlines. That this is also the case for the ratios of their IAs indicates that despite the different physical conditions along these sightlines, the carriers of the two types of bands are equally resilient to or equally vulnerable to them. This behavior is consistent with expectations for structural end members (see above). 

Broader Galactic trends in dust extinction and composition provide additional perspective on the relative abundances observed along the Cyg OB2-12 and Quintuplet sightlines.
 \cite{Roche85} found that on sightlines to a central cluster of bright stars, located within 2 pc of Sgr A*, the strength of the 9.7 \mum\ silicate feature relative to the near-IR continuum extinction, e.g., $A_K$, is approximately twice its value toward Cyg OB2-12.  From the silicate values (see bottom two rows of Table~\ref{tab:t1}) and estimates of $A_K$\footnote{We use a $K$-band extinction toward Cyg OB2-12 of $1.00 \pm 0.15$ mag, based on \citet{rieke85}, \citet{hanson2003}, and \citet{wright2015massive}.
 For the Quintuplet we adopt the value $A_K$ = 3.1 mag, the average of its values in \citet{figer1999massive} and \citet{liermann2010quintuplet}.}, the ratio toward the Quintuplet, located 30 pc distant from Sgr A*, is $\sim$ 1.4 times higher than it is toward Cyg OB2-12 (which is considered to be in the solar neighborhood). Therefore, while not as extreme as the central cluster, the Quintuplet still shows a significant silicate enhancement over the Cyg OB2-12 baseline. 

 The ratios IA(3.3)/$A_K$ and IA(3.4)/$A_K$ (where $A_K$ denotes $K$-band extinction) are similar for the two sightlines, as expected if the hydrocarbon absorption bands and the $K$-band extinction arise from the same carbonaceous dust population \citep{Draine84}.
 Similar values on both sightlines are also seen in the mid-IR ratio, IA(6.2)/$A_K$.  Although the carbon and silicon abundance (gas plus dust) increases toward the Galactic center, as indicated by chemical evolution models \citep{hou2000} and supported by observational studies of metallicity gradients across the Galactic disk \citep{rolleston2000obstars,andrievsky2002cepheids}, this does not necessarily result in equal increases in the silicate and  carbonaceous stardust injected into the ISM. The increased metallicity in the inner Galaxy requires that AGB stars have to dredge up more carbon from their cores to change from O-rich to C-rich and, hence, overall the production of carbon stardust relative to silicate dust may well decrease \citep{Roche85}.

\subsection{Interstellar Dust Masses and Volumes}
\label{sec:dust mass}
Following \cite{tielens1987composition}, we have quantified the physical contribution of various dust components by converting column densities\footnote{These column densities were derived from the integrated areas of the observed absorption bands using laboratory-determined band strengths, as described in Section~\ref{sec:dust variations}, and are listed in Table~\ref{tab:t1}.} derived from absorption band strengths into corresponding dust masses and volumes per hydrogen atom for the Cyg OB2-12 and Quintuplet sightlines. Specifically, 
\begin{equation}
    m_{d,i}\, =\, \frac{N_im_im_H}{N_H},
\end{equation}
with $N_i$, $m_i$, $m_H$, $N_H$ the column density of the specific atom/group in dust component $i$ (Table~\ref{tab:t1}), the combined atomic weight in amu of this atom/group, the mass of hydrogen, and the column density of H along the line of sight, respectively. We have adopted a mean atomic weight per Si atom of 172 amu for the silicates (corresponding to MgFeSiO$_4$), and an H column density of 1.9~$\times$~10$^{22}$~cm$^{-2}$ for the Cyg OB2-12 sightline, corresponding to 10 magnitudes of visual extinction in the diffuse ISM \citep{savage1977}. For the Quintuplet, we assume a diffuse ISM extinction of 20 mag (see Section~\ref{sec:Q_LOS}), corresponding to $N_H$ = 3.8 $\times 10^{22}$~cm$^{-2}$. Group masses were based on atomic weights: 13 amu for CH, 14 amu for CH$_2$, 15 amu for CH$_{3}$, and 12 amu for each aromatic or olefinic carbon. Mass and volume ratios relative to silicates were obtained by dividing the mass or volume per H atom for each carbonaceous component by that of the silicate component along the same sightline. For the Quintuplet, only these relative ratios are reported due to greater uncertainty in the absolute hydrogen column. Column densities for the hydrocarbon features were scaled to a diffuse visual extinction of 20~mag (see Section~\ref{sec:Q_LOS}), consistent with the assumption that hydrocarbon absorption arises entirely in the diffuse ISM.
These masses have been converted into dust volumes per H-atom, $V_i=m_{d,i}/\rho_i$ adopting specific densities, $\rho_i$, for aromatic carbon, aliphatic carbon, and silicates of 2.2, 1.8, and 3.0 g cm$^{-3}$, respectively \citep{tielens1987composition}. For olefinic carbon, we have used the measured specific density of ethylene, 0.65 g/cm$^{-3}$ \citep{satorre2017}.

From Table \ref{tab:t1}, the total column density of solid carbon on the Cyg OB2-12 sightline, including the estimates from the aromatic and olefinic CC modes and accounting for aliphatic carbon through the aliphatic CH stretching modes of CH$_2$ and CH$_3$ groups\footnote{This total is the sum of the carbon column densities listed in Table~\ref{tab:t1} for the Cyg OB2-12 sightline: $3.6 \times 10^{18}$~cm$^{-2}$ from the 6.25 \mum\ aromatic CC mode, $0.77 \times 10^{18}$~cm$^{-2}$ from the 6.05 \mum\ olefinic CC mode, and $0.18 \times 10^{18}$ from the sum of the CH$_2$ and CH$_3$ groups of the 3.4 \mum\ aliphatic CH stretch.}, is 4.5 $\times$ 10$^{18}$ cm$^{-2}$. The total H column density (see above) then results in an abundance of carbon in dust of 2.4 $\times$ 10$^{-4}$ along this sightline.\footnote{Computed as $N(\mathrm{C}) / N(\mathrm{H}) = (4.5 \times 10^{18}) / (1.9 \times 10^{22}) = 2.4 \times 10^{-4}$. }
 The gas-phase carbon abundance toward Cyg OB2-12 is not known, but typically is 1.6$\times$ 10$^{-4}$ in the interstellar medium \citep{sofia2004}. This results in a total carbon abundance of 3.9 $\times$ 10$^{-4}$. 
 
 The solar carbon abundance is 2.7 $\times$ 10$^{-4}$ \citep{Asplund2009}.  However, solar elemental abundances may not be representative of the interstellar abundances because of settling of heavy elements in the solar photosphere \citep{Lodders2003} and because continued chemical enrichment by stars may have increased the interstellar carbon abundance over the last 4.5 Gyr \citep{Rybizki2017}. In that respect, abundances of young F and G stars may be more representative of current interstellar abundances. The measured carbon abundance in young F and G stars of 3.6 $\times$ 10$^{-4}$ \citep{Sofia2001} is indeed in much better agreement with our estimate of 3.9 $\times$ 10$^{-4}$ for the total solid- and gas-phase carbon abundance. 
 
The inferred Si abundance in silicate dust is 5.25 $\times$ 10$^{-5}$ (c.f., Table~\ref{tab:t1}) toward Cyg OB2-12. For comparison, the F and G star abundance of silicon is 4.0 $\times$ 10$^{-5}$. As the intrinsic strength of the 10 \mum\ SiO stretching band varies by about 30\%\ among different silicate compounds \citep{tielens1987composition}, this discrepancy may not present a real issue. 

 Similar mass and volume estimates for the Quintuplet sightline are included in Table~\ref{tab:dust mass}. Because the hydrocarbon features are formed only in the diffuse ISM component of that sightline, we assume a visual extinction of 20~mag and scale the column densities accordingly (see Section~\ref{sec:Q_LOS}). The resulting hydrocarbon-to-silicate mass and volume ratios are roughly comparable to those for Cyg OB2-12, suggesting that the diffuse ISM along both sightlines contains a similar mix of carbonaceous dust components.

\begin{table*}[!hbt]
\fontsize{8.5}{8.5}\selectfont
\caption{\label{tab:dust mass} Relative Masses and Volumes of Carbonaceous Compounds and Silicates
toward Cyg OB2-12\tablenotemark{a} and the Quintuplet\tablenotemark{a,b}} 
\begin{tabular}{|l|c|c|c|c|}
\toprule
Sightline and Mode & Mass /H atom  & Mass/Silicate Mass\tablenotemark{c} & Volume/H atom & Volume/Silicate Volume\tablenotemark{c} \\
 & $10^{-27}$ g & &$10^{-27}$ cm$^{3}$ & \\
 \hline\hline
Cyg OB2-12 Aromatic\tablenotemark{d,e} & $3.8 \pm 0.2$\tablenotemark{e} & $0.26\pm 0.02$  & $1.7\, \pm 0.1$ & $0.35 \pm 0.02$ \\
Quintuplet Aromatic & & $0.17 \pm 0.01$ &  & $0.34 \pm 0.02$ \\
\hline
Cyg OB2-12 Aliphatic (HAC) \tablenotemark{f} & $0.23\, \pm 0.05$& $0.015 \pm 0.004$ & $0.12\, \pm 0.03$ & $0.024 \pm 0.05$ \\
Quintuplet Aliphatic (HAC) \tablenotemark{f}  & & $0.011 \pm 0.002$ &  & $0.018 \pm 0.004$\\
\hline
Cyg OB2-12 Olefinic & $0.81\, \pm 0.08$ & $0.054 \pm 0.006$ & $1.21\, \pm 0.14$ & $0.25\pm0.03$ \\
Quintuplet Olefinic & & $0.081 \pm 0.009$ &  & $0.38 \pm0.04$ \\
\hline
Cyg OB2-12 Silicates & $14.9\, \pm 0.3$ & & $4.9\, \pm 1.0$ & \\
\hline\hline
\end{tabular}

\tablenotetext{a}{We adopt a hydrogen column density per unit extinction of $N(\mathrm{H})/A_V = 1.9 \times 10^{21}~\mathrm{cm}^{-2}~\mathrm{mag}^{-1}$ for both sightlines \citep{savage1977}. For Cyg OB2-12, we use $A_V = 10$~mag; for the Quintuplet, we use $A_V = 20$~mag (assuming bands are formed only in the diffuse ISM, which contains two-thirds of the hydrogen and extinction; see Section~\ref{sec:Q_LOS}). Specific densities of each dust component are given in Section \ref{sec:dust mass}.}
\tablenotetext{b}{For the Quintuplet sightline, only mass and volume ratios relative to silicates are reported. These were calculated using hydrocarbon column densities scaled to 20~mag of diffuse extinction (see Section~\ref{sec:Q_LOS}). Silicate values were assumed to scale proportionally from those used for Cyg OB2-12, consistent with the factor of two increase in diffuse $A_V$.}
\tablenotetext{c}{Mass and volume ratios are calculated by dividing each component’s mass or volume per H atom by the corresponding silicate value along the same sightline. For example, the aromatic component toward Cyg OB2-12 has a mass to silicate mass ratio of $(3.8 \times 10^{-27}) / (14.9 \times 10^{-27}) = 0.26$, and a volume to silicate volume ratio of $(1.7 \times 10^{-27}) / (4.9 \times 10^{-27}) = 0.35$.}
\tablenotetext{d}{Uses column density of CC {\it sp}$^{2}$~6.25 \mum\ aromatic band in Table \ref{tab:t1}.}
\tablenotetext{e}{As an example, from Equation 3 the mass per H atom is $3.6 \times 10^{18} \times 12 \times 1.66 \times 10^{-24}~\mathrm{g} \,/\, 1.9 \times 10^{22} = 3.8 \times 10^{-27}$~g.}
\tablenotetext{f}{HAC is assumed to contain the aliphatic hydrocarbon component. In Equation 3, for both sightlines we use a mean atomic mass of 14.4 amu for the CH$_{2}$ and CH$_{3}$ groups that combine to produce the 3.4 \mum\ band.}
\end{table*}

These estimated masses and volumes offer a physically intuitive view of the relative contributions of each dust component. They support the conclusion that carbon-rich material is a significant component of diffuse interstellar dust. The results show that aromatic carbon dominates both the mass and volume budget of the carbonaceous material along both sightlines, followed by olefinic and aliphatic components. The mass and volume ratios relative to silicates indicate that while carbonaceous compounds represent a substantial fraction of the total dust population, their contribution is consistently lower than that of silicates.

\section{Discussion: the structure of interstellar carbon dust}
\label{sec:Disc} 

We begin this section by summarizing the infrared absorption features and associated chemical structures that characterize interstellar carbonaceous dust. In light of the striking spectral similarities observed toward Cyg OB2-12 and the Quintuplet, two widely disparate sightlines, we then examine what these data reveal about the structure and evolution of carbonaceous grains in the diffuse interstellar medium.

\subsection{Infrared Features and Chemical Composition}
\label{sec:Obs_constraints}
  Based largely on infrared observations, it is generally accepted that dust in the diffuse ISM contains carbonaceous and silicate grains. Polarization studies toward the Galactic center reveal that the 10 \mum\ silicate feature is polarized while the 3.4 \mum\ hydrocarbon dust feature is not, implying that silicate and hydrocarbon dust features originate in two distinct dust components rather than as hydrocarbon dust mantles on top of silicate cores \citep{adamson1999spectropolarimetric,chiar2006spectropolarimetry}. The carbonaceous component includes both large PAHs ($\simeq$30-100 carbon atoms) and submicron-sized carbonaceous grains. The 2175 \AA\ feature is due to the $\pi\rightarrow \pi^\star$ electronic transition in aromatic carbon structures with typical sizes of 200 \AA\ ($\simeq$5$\times$10$^6$ C-atoms), but the carbon grain size distribution might well extend to $\simeq$3000 \AA\ \citep[$\simeq$ 5$\times$ 10$^9$ C atoms,][]{Mathis77,jones2012a, jones2012b, tielens2005physics}. They encompass a range of bonding types including single bonds (as in alkanes), double bonds (as in alkenes), 
  and aromatic carbon in ring structures (e.g., benzene-like). There is also infrared evidence presented in this paper for the presence of oxygen-bearing functional groups, such as carbonyl (C=O) and hydroxyl (-OH), associated with the carbonaceous material. 

As shown earlier in this paper, the absorption features that we detect in the near- and mid-IR spectra of the Cyg OB2-12 and Quintuplet sightlines can be accurately decomposed into Gaussian components. The components responsible for the 3.4 \mum\ absorptions were identified in earlier studies \citep[e.g.,][]{sandford1991interstellar,pendleton1994near,whittet1997infrared,rawlings2003infrared,chiar2013structure}. For the mid-IR spectra, some additional comments are in order. Both the 5.85 \mum\ and 6.05 \mum\ bands can be attributed to the carbonyl (C=O) stretching mode, with the former originating from a carbonyl attached to an aliphatic component and the latter involving two carbonyl groups attached to a single aromatic unit (i.e., a quinone-like structure) of the carbonaceous material, where the double bonds are arranged to maintain conjugation  \citep{socrates}.  Although quinones are a plausible carrier given the aromatic structure of the carbonaceous dust, this identification is not definitive, as other carbonyl-containing molecules, including unsaturated conjugated ketones  (structures with carbonyl groups linked to double bonds) or cyclic compounds, could also contribute to this feature \citep{socrates}. Here, we consider the 6.05~\mum\ band as arising from quinones.

\subsection{Origin and Evolution of Interstellar Carbon Dust }
\label{sec:origin_evolution}

\subsubsection{Structural Models}
\label{structuralmodels}

As described by \cite{robertson2002}, HAC consists of small sp$^{2}$ carbon clusters embedded in a disordered tetra-coordinated “matrix” formed by the {\it sp}$^{3}$ carbon atoms. These aromatic units are closely linked in a network that lacks long-range order, with aliphatic and olefinic structures bridging the clusters \citep{robertson2002,jones2012a}. This structure is distinct from that of soot produced in hydrocarbon flames, which are assemblies of “basic structural units” consisting of of aromatic layers akin to clusters of PAH-like species, possibly very loosely connected by aliphatic bridges \citep{russo2013probing,jaeger2009A}. 

In the ISM, both types of carbonaceous material appear to be present: HAC-like material through the 3.4~\mum\ and 6.85~\mum\ absorption features as well as the olefinic bands, while soot-like structures are traced by the  3.3~\mum\ and 6.2~\mum\ absorption bands and the PAH emission features. We emphasize that the detailed profile of the 3.4 \mum\ band implies that the CH$_2$ and CH$_3$ groups are not intermixed into large aromatic structures \citep{ristein1998comparative} and the HAC and aromatic features arise from two distinct interstellar dust components. This interpretation aligns with the work of \citet{leger1989}, who proposed that, in emission, the  3.4~\mum\ band arises from CH$_2$/CH$_3$ groups attached to PAHs, whereas in absorption the 3.4~\mum\ absorption feature observed in the diffuse ISM is best matched by HAC materials.

The presence of the 3.4~\mum\ absorption band in the diffuse interstellar medium, together with its near-total absence in the ejecta of carbon-rich stars responsible for most stardust \citep{chiar2013structure}, strongly suggests that its carrier forms through processing in the ISM. 

\subsubsection{Carbon Dust Core-Mantle Structure} 
\label{core-mantle}

Several models have been proposed for the structure of interstellar carbon dust. Some interpretations consider all carbonaceous units as part of one amorphous structure (a-C:H) \citep{Jones2017}. In contrast, laboratory studies suggest that aliphatic material may form as a surface layer atop aromatic grains through hydrogenation processes in the ISM \citep{mennella2002ch, chiar2013structure}.  The general chemical composition inferred from our observations is consistent with earlier interpretations based on lower-resolution spectra \citep{pendleton2002organic}. The more specific 
core–mantle structure, with aromatic cores coated by thin HAC-like mantles, was first introduced by \citet{chiar2013structure}. As discussed below, analysis of the spectra presented here provides new observational support for this core–mantle picture, showing that aliphatics and carbonyls are likely surface features on primarily aromatic grains.

\subsubsection{Aliphatic and Carbonyl Group Abundances}
\label{groupabundances}

The CH$_2$ column densities in Table~\ref{tab:t1}, derived from both the CH stretching  (3.4~\mum) and CH$_2$ scissoring  (6.85~\mum) bands toward Cyg OB2-12  are in reasonable agreement, lending confidence to the derived abundance of aliphatics.
Although a quantitative estimate for the CH$_3$ content cannot be derived from the deformation modes in the 5$-$8~\mum\ range, we can conclude from the values for the stretching mode (i.e., the 3.3~\mum\ and 3.4~\mum\ data in Table~\ref{tab:t1}) that the column densities of CH$_2$ groups somewhat exceed those of the CH$_3$ groups. 
Oxidation of the interstellar aliphatic carbon dust component seems to be of limited importance, based on the relatively weak C=O 5.85~\mum\ band in Table~\ref{tab:t1}, which gives an O/H ratio of $\simeq$1/20. The same conclusion can be drawn from the similar depth of the shoulder at 6.05~\mum\ attributed, possibly, to the C=O quinone stretch.  

\subsubsection{Spectral Band Assignments and Structural Implications}
\label{structuralimplications}

Following \citet{chiar2013structure}, the 6.19 and 6.25~\mum\ bands are attributed to C=C stretching modes in olefinic and aromatic structures, respectively, and thus also probe the aromatic component of the dust. In contrast, the 6.85 \mum\ band is attributed to the CH$_2$ scissoring mode in saturated aliphatic compounds \citep{ristein1998comparative} and like the aliphatic CH stretching modes near 3.4~\mum, is assigned to the aliphatic dust component. In principle, the CH$_3$ asymmetric deformation mode, which absorbs near 6.85~\mum, could contribute to the observed absorption band at 6.85~\mum. However, the corresponding CH$_3$ symmetric deformation mode at 7.27~\mum, which is typically stronger than the 6.85 \mum\ band in laboratory spectra \citep{ristein1998comparative,wexler1967integrated}, appears to be weak or absent in the Cyg OB2-12 spectrum (see Figure~\ref{fig:fig2}). This argues against a strong CH$_3$ bending component and supports our attribution of the interstellar 6.85~\mum\ band to the CH$_2$ scissoring mode, a conclusion consistent with the locations of aliphatic chains in a connected HAC network. 

\subsubsection{Hydrogenation at the Atomic Scale: H/C Ratios}
\label{H/CRatios}
The H/C ratio of the interstellar aromatic carbon dust component, based on the aromatic 3.3~\mum\ CH and aromatic 6.25~\mum\ CC column density values in Table~1, is 
$\simeq$0.13 for each sightline. 
This ratio includes only the hydrogen atoms directly bonded to aromatic carbon atoms and it serves as a tracer of aromatic hydrogenation.
A ratio of $\sim$0.13 suggests moderately large aromatic domains with a relatively small number of hydrogen atoms at their edges. For compact PAHs, the H/C ratio is given by $x^{-1}$ with the structure given by  3x$^2-$3x+1 hexagonal (benzene) cycles arranged in x-1 rings around a central hexagon \citep{tielens2005physics}. The observed H/C ratio for aromatic dust corresponds to x=7.5 and the chemical formula, C$_{6x^2}$H$_{6x}$=C$_{340}$H$_{45}$. Such a compact species would have a size of 0.9~$\times N_C^{1/2}\simeq$20 \AA\ radius. Although the IR spectra demonstrate that the HAC component traced by the 3.4~\mum\ band is not related to the aromatic component, in principle, the observed olefinic component could be intermixed with the aromatic dust component as bridges between aromatic units, leading to an ``effective'' H/C of 0.24 and an aromatic structure of C$_{96}$H$_{24}$(CC)$_{19}$\footnote{That is, 96 aromatic C-atoms in hexagons terminated by 24 aromatic H's and 19 olefinic CC bridges.}. However, we consider it more likely that the olefinic component is intermixed at the molecular level with the HAC dust.

 The presence of oxygen does not appear to significantly influence these size estimates, as the intrinsic strengths of oxygen-related features, such as the C=O stretch in carbonyl groups (5.8$-$6.1~\mum), and the very broad OH stretch near 3.1~\mum—are typically 2--20 times greater than that of the 3.4~\mum\ aliphatic CH stretch band \citep[e.g.,][]{wexler1967integrated, dhendecourt1986, pendleton2002organic}. The relative weakness of oxygen-bearing features in the observed spectra therefore implies that oxygen is only a minor component of the carbonaceous dust.

The estimates for molecular composition and domain size can be directly compared to known PAHs and soot-like materials which span a similar range of structures. For example, common PAHs such as coronene (C$_{24}$H$_{12}$), circumcoronene (C$_{54}$H$_{18}$), and circumcircumcoronene (C$_{96}$H$_{24}$) span the typical size range of interstellar PAHs \citep{Croiset2016}. 
Laboratory experiments of soot particles show aromatic domains extending up to C$_{250}$, with a peak around C$_{40}$ \citep{jaeger2009A}.

\subsubsection{Dust Formation and Processing in the ISM}
\label{dustformation}

The aromatic dust component in the ISM is thought to have formed in the outflows from C-rich AGB stars in chemical processes analogous to soot formation in terrestrial flames \citep{frenklach1989,cherchneff1992}. In these environments, the nucleation and growth processes begin with C$_2$H$_2$ as a feedstock, leading to the formation of large PAH species. These PAHs then aggregate into clusters, which then, in turn coagulate into soot particles. Laboratory studies show that this soot consists of a broad range of PAH monomer sizes -- from C$_{20}$H$_x$ to C$_{250}$H$_y$ \citep{jaeger2009A}. There is a high temperature soot formation window which produces fullerene-like structures \citep{jaeger2008}, but those high temperatures are not relevant for outflows from AGB stars.

As pointed out earlier most aliphatic dust is thought to form in the diffuse ISM; the mechanism is believed to be processing of aromatic carbon dust there.  Specifically, thermal hydrogen atoms interact with aromatic surfaces to produce HAC material. This process is counteracted by 
ultra-violet photons that drive photo-bleaching (photon-driven removal of H) in the strong interstellar far-UV  radiation field \citep{mennella2001uv,mennella2006synthesis}. The structure of the resulting surface layer is a balance between these two processes. Simple models and laboratory studies suggest that when aromatic carbon dust is exposed to hydrogen atoms in the diffuse ISM, a thin surface layer of hydrogenated material quickly forms that reaches ``saturation'' after roughly 10$^4$ years, at which point hydrogen atoms occupy all available bonding sites in this surface layer.

This process is halted inside dense clouds because the formation of ice mantles prevents  H-addition, while cosmic ray processing will continue to drive off H from the HAC mantle underneath the ice \citep{mennella2001uv}. We note that this energetic processing may also remove some of the aromatic H, converting the PAH-like character of the basic structural units making up aromatic grains into graphene-like flakes.

If HAC mantles saturate at a fixed thickness set by the typical depth, $d$,  at which UV photons are absorbed in a dust grain (Beer's law gives, $d \sim \lambda/4\pi\simeq$ 100 \AA ), then the overall C/H ratio of carbonaceous dust will not be strongly affected by local physical conditions. However, the ratio of aromatic to aliphatic absorption should vary with grain size, since smaller grains have a higher surface-to-volume ratio and thus a proportionally thicker hydrogenated layer.
 The remarkable similarity of  the aromatic-to-aliphatic ratios toward Cyg OB2-12 and the Quintuplets (as evidenced by IA(6.25)/IA(3.4) and IA(3.3)/IA(3.4) which trace the relative contributions of aromatic C=C and CH modes to aliphatic CH modes, respectively) suggests then that the  grain sizes along these two sightlines are comparable, and that their HAC mantles have reached similar degrees of hydrogenation.

 \begin{figure*}[!htb]
\epsscale{1.2}
\includegraphics[clip, trim=0cm 0cm 0cm 0cm, width=\linewidth]
{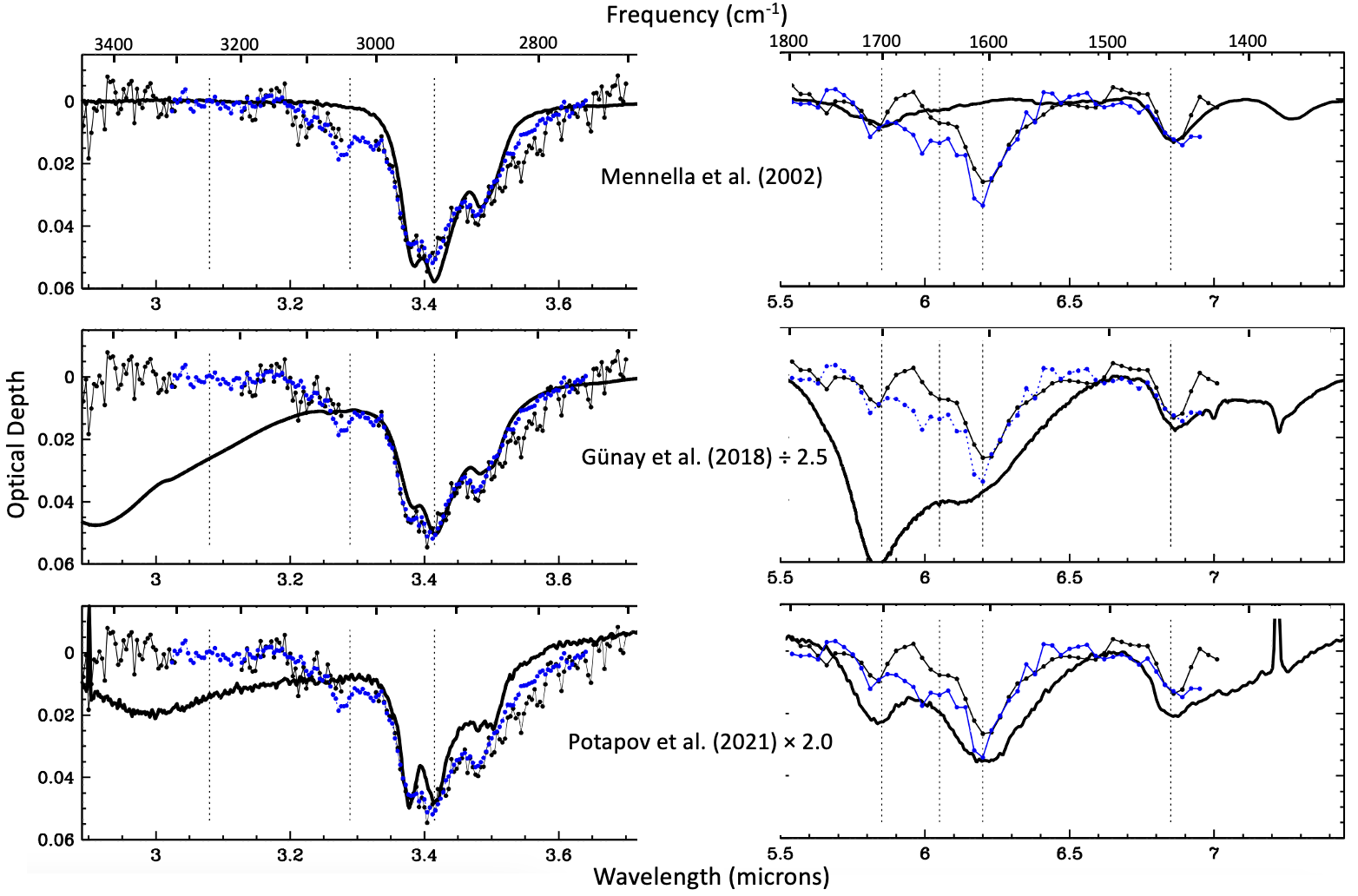}
\caption{\label{fig:fig7}Optical depth near-IR and mid-IR spectra for Cygnus OB2-12 (black dots) and Quintuplet (blue dots) compared to three laboratory spectra (black lines). Wavelength scales are at the bottom of each spectrum; frequency scale is at the top of the figure. Scaling factors shown apply to both near-IR and mid-IR data. Dashed vertical lines are at 3.08~\mum\ (corresponding to water ice), 3.289~\mum\ (the central wavelength of the Gaussian used to fit the 3.3~\mum\ CH aromatic band), 3.415~\mum\ (the 3.4~\mum\ aliphatic band), 5.85~\mum\ (aliphatic carbonyl band), 6.05~\mum\ (aromatic, quinone carbonyl band), 6.2~\mum\ (blended olefinic and aromatic bands), and 6.85~\mum\ (scissor band). See Table~1 for details. Top \citep{mennella2002ch}: Hydrogenated amorphous carbon (HAC) produced by 
hydrogen atom irradiation of carbon grains under simulated diffuse interstellar medium conditions, resulting in predominantly surface aliphatic hydrocarbons. Middle \citep{gunay2018}: An interstellar dust analogue synthesized from a carbon-rich plasma using an acetylene precursor under simulated circumstellar conditions, yielding predominantly graphitic particles with an aliphatic hydrocarbon component. Bottom  \citep{potapov2021dust}: UV-irradiated H$_2$O:CH$_3$OH:silicate mixture in which the components were deposited together at low temperature and then exposed to ultraviolet light. This mid-IR spectrum has been flattened by dividing it by a linear continuum matching the spectrum at 5.55~\mum\ and 7.35~\mum. The three laboratory analogs were specifically prepared to reproduce the 3.4~\mum\ aliphatic CH stretch complex. Differences between the laboratory spectra and the interstellar absorption at other wavelengths (e.g., the 3.3~\mum\ aromatic CH stretch or carbonyl bands near 5.85~\mum) are expected, as these experiments did not attempt to replicate the full chemical complexity of interstellar carbonaceous dust.}
\end{figure*}

\subsection{Formation Scenarios Informed by Laboratory Analogs} 
\label{sec:formation}

Several detailed laboratory studies have been performed to address the origin of the aliphatic signatures of HAC grains (at 3.4 and 6.8~\mum) and their implications for the origin and evolution of interstellar carbon dust. We emphasize that these studies were not intended to address the infrared absorption features associated with the aromatic carbon material in the interstellar spectra (at 3.3 and 6.2~\mum). Hence, in discussing the comparisons shown in 
Fig.~\ref{fig:fig7},
we will not take the latter features into account.

Comparisons of the interstellar absorption features with laboratory spectra from 
\citet{mennella2002ch}, \citet{gunay2018}, and \cite{potapov2021dust}, shown in 
 Figure~\ref{fig:fig7},illustrate the success of analog materials in reproducing the aliphatic CH stretch complex near 3.4~\mum\ and the aliphatic 6.85~\mum\ band on the Cyg OB2-12 and Quintuplet sightlines. The laboratory spectra successfully reproduce both the detailed profile of the 3.4~\mum\ absorption complex and the relative intensities of the 3.4~\mum\ and 6.85~\mum\ bands, closely matching astronomical observations in those regions. We note that \citet{schnaiter1999} also produced an analog that successfully reproduces the aliphatic bands, as shown and discussed in \citet{pendleton2002organic}.

These laboratory analog studies provide supporting evidence for the structural picture inferred from our observational results. The analog materials shown, including HAC analogs and UV-processed organic residues, were specifically prepared to reproduce the aliphatic features and successfully match the relative strengths and profiles of the methyl and methylene bands that contribute to the 3.4 \mum\ features.Differences between the laboratory spectra and the interstellar absorption at other wavelengths (e.g., the 3.3~\mum\ aromatic CH stretch or carbonyl bands near 5.85~\mum) are expected, as these experiments did not attempt to replicate the full chemical complexity of interstellar carbonaceous dust. 

Beyond their ability to reproduce the observed aliphatic absorption bands, these laboratory analogs also offer insights into possible formation and processing pathways for interstellar carbon dust. We now describe  the three laboratory analogs presented from a different perspective, focusing on the formation and processing scenarios they represent rather than their abilities to reproduce interstellar spectra.  Each of them refers to a different scenario for the origin and evolution of carbon dust in the ISM.

 The HAC model of  \cite{mennella2002ch}  envisions carbon-rich stardust injected by C-rich AGB stars and processed by thermal hydrogen and UV-bleaching in the ISM (e.g., \citep{pendleton2002organic}, \cite{chiar2013structure}).
In this way, a thin HAC layer is produced on and bleached off of an aromatic carbon dust core. Thus, the HAC and aromatic compounds are intimately related, naturally explaining the absence of the 3.4 \mum\ band inside dense clouds, because the development of the water ice mantle on dense cloud grains inhibits reforming the aliphatics in the dense clouds \cite{mennella2008activation}.
 
The plasma study by \citet{gunay2018} is, in some ways, a variant of this model where the HAC component is already produced during the plasma deposition step rather than by subsequent processing in the ISM. The absence of the 3.4 \mum\ feature in the prototypical C-rich AGB star, IRC+10216, argues against this. It is possible that conditions in other C-rich dust outflows might be more conducive to HAC formation, which would suggest that the HAC and aromatic dust components are carried by separate dust components \citep{jones2011dust}.

In contrast, \citet{potapov2021dust} propose that carbonaceous dust forms via UV photolysis of interstellar ices deposited on silicate grains -- an idea rooted in earlier work by \citet{greenberg1995}. Although this process can produce complex organic residues in the lab, it faces challenges under astrophysical conditions: UV fluences may be insufficient in the ISM, and typical grain temperatures are too low to fully drive off the volatile ice matrix \citep{planck-dust2014}. Observationally, such a model also faces a challenge in explaining the absence of the 3.4~\mum\ band in dense clouds. In addition, the interstellar 9.7~\mum\ silicate and 3.4~\mum\ carbon dust features do not show the same polarization behavior and hence are not intimately linked as core-mantle grains \citep{adamson1999spectropolarimetric,chiar2006spectropolarimetry}. 

 Further experimental studies and more detailed astronomical models may be able to address some of these challenges. Taken together, the observational constraints, laboratory analog comparisons, and formation pathway considerations all point to a model in which HAC forms thin surface layers on pre-existing aromatic carbon grains in the diffuse ISM \citep{chiar2013structure}.

\subsection{Interstellar dust models}

While a full-fledged analysis of the implications of this study for interstellar dust models is beyond the scope of this paper, it is of some interest to briefly consider the implications in the context of such models. Our discussion of the data presented here as well as the discussion in \citet{chiar2013structure} recognize the presence of three distinct dust compounds – silicates, aromatic carbon, and HAC – and the need for them to be accommodated in models of the composition, origin and evolution of interstellar dust.  Early models relied on the injection of separate graphitic and silicate dust components by C-rich and O-rich Asymptotic Giants Branch stars \citep[]{Mathis77, 1994ApJ...422..164K, Draine84, 2004ApJS..152..211Z}. Later models explicitly included HAC to accommodate the presence of the 3.4 \mum\ absorption band \citep{jones2012a,jones2012b}. 

The results of the Planck mission have driven a further evolution of these global models for interstellar dust. In particular, Planck revealed little variation in the degree of polarization of dust emission in the sub-millimeter ($>300$ \mum) regime \citep{2015A&A...576A.107P, 2016A&A...586A.132P}. 
This finding poses challenges for early models based on separate dust components, as the carbonaceous grains were not aligned and the alignments of silicate and graphite emission vary differently with wavelength 
\citep{2009ApJ...696....1D}. Some models accommodate this by combining carbonaceous and silicate dust into large fluffy aggregates \citep{2023ApJ...948...55H},
presumably formed through coagulation in dense environments 
\citep[c.f.,][]{2009A&A...502..845O}. Other interstellar dust models have assumed that both large silicate and large HAC-bearing grains can be aligned \citep{2024A&A...684A..34Y}. We emphasize that, in agreement with the conclusions of the \citet{chiar2013structure} study and the results presented here, both of these models assume that the HAC component is a thin layer covering large interstellar dust grains. 

The observed difference in polarization of the 3.4 and 10 \mum\ features has long been a point of concern for interstellar dust models \citep{chiar2006spectropolarimetry}. As emphasized by 
\citet{2021ApJ...909...94D}, this difference might be explained if the HAC component is a thin surface layer whose thickness is independent of core size and if small grains are not aligned. Alternatively, it may be that models developed for dust in the diffuse ISM are not applicable for the denser environments probed by these spectropolarimetry measurements \citep{2024A&A...684A..34Y}. 
 
 Throughout this study, the evolution of carbon dust and the formation of HAC material in the diffuse ISM have been placed in the context of reactions of thermal H atoms with carbonaceous grains forming HAC and the UV-bleaching of the resulting HAC mantles. While these processes have been well studied in the laboratory \citep[e.g.][]{mennella2001uv, mennella2002ch, mennella2003effects}, other interpretations are possible. Specifically, recent studies of carbon abundances in diffuse sightlines, based on the strong 1334 \AA\ transition of C$^+$, imply that carbon depletion is highly variable.\footnote{Note that early studies of C depletion -- based on the weak 2325 \AA\ transition -- led to the conclusion that there is little variation in the C depletion in the diffuse ISM. This discrepancy has not yet been resolved (\cite{2011AJ....141...22S} and \cite{parvathi2012probing}). } Such variability is indicative of rapid interchange between the gas and solid phases of the ISM, presumably reflecting sputtering in interstellar shocks and chemical reactions in diffuse clouds. Although the chemistry involved has not been studied, it seems possible that the highly reducing environment of the ISM might lead to the formation of HAC mantles. This would also lead to the formation of thin HAC layers on top of dust cores, but we note that in this case, it is conceivable that silicate grains also acquire HAC mantles \citep{2024A&A...684A..34Y}. The absence of HAC material in molecular clouds would, however, still require a local bleaching process. In passing, we note that such mantles might serve as thin ``veneers'' that protect the underlying grains from the cumulative effects of sputtering in shocks \citep{1998ApJ...499..267T,  2009A&A...500..335T} . 
This may alleviate the discordance between the short lifetime estimates for interstellar dust ($\simeq 500$ Myr; \cite{jones1994grain} and the very long stardust injection timescale (2.5 Gyr; \cite{tielens2005physics} .

As this short discussion indicates, it is clear that further studies are warranted. These should include observational studies of a large sample of carbonaceous and silicate infrared absorption features and their dependences on column density as well as their behaviors with depth in molecular clouds. They should also include studies of the dependences of feature properties on elemental depletions on diverse sightlines, and infrared polarization measurements along sightlines where optical polarization has also been measured. On the laboratory side, further studies of the low temperature chemistry involving silicon, carbon, and oxygen in reducing atmospheres are needed. On the theoretical side, modeling the alignment of carbonaceous materials as well as the stability of loosely bound grain aggregates under diffuse cloud and warm intercloud conditions would be highly relevant.

\section{Conclusions}
\label{sec:Summary}
 We present a new 2.86$-$3.70~\mum\ spectrum of the highly extinguished hypergiant Cyg OB2-12, a key sightline for studying interstellar dust chemistry without contamination from dense molecular cloud material. This spectrum, obtained at higher resolution than previous spectra, enables detailed comparisons of the near-IR hydrocarbon absorption bands in Cyg OB2-12 to those observed toward the more heavily extinguished Quintuplet star cluster in the GC. We pair this analysis with an examination of the 5.5$-$7.0~\mum\ portion of a previously published (but here re-processed) \textit{Spitzer} spectrum of Cyg OB2-12 and an ISO SWS spectrum of the Quintuplet in the same wavelength interval. Together, these data provide the first opportunity to compare the Cyg OB2-12 and Quintuplet sightlines in sufficient detail to evaluate the composition of carbonaceous dust in different interstellar environments.

Although the sightlines to Cyg OB2-12 and the Quintuplet differ greatly in their Galactic orientations, lengths, and compositions of interstellar medium, they exhibit remarkably similar absorption features associated with hydrocarbons. Spectra of both contain features at 3.3, 3.4, 6.2, and 6.85~\mum\ that reveal the presence of aromatic (3.3 and 6.25~\mum) and aliphatic (3.4 and 6.85~\mum) hydrocarbons, along with carbonyl-bearing compounds at 5.85 and 6.05~\mum, and olefinic material near 6.19~\mum. The relative strengths of the hydrocarbon bands along each sightline indicate that the carbonaceous dust along each is predominantly aromatic and broadly similar in composition. 

Moreover, the near equality of the ratios of the integrated areas of aromatic bands to aliphatic bands on each sightline indicates that whatever the clearly different physical conditions are on the diffuse parts of these sightlines and portions thereof, the carriers of the two types of bands are equally resilient to them and equally vulnerable to destruction due to them.

We present an argument, consistent with previous conclusions from numerous studies, that aliphatic hydrocarbons are confined to diffuse interstellar regions. This implies that they are only present in diffuse portions of the Quintuplet sightline, not in denser spiral arm material along that sightline. The argument also applies to the aromatic 3.3~\mum\ bands (and presumably the 6.25~\mum\ bands) seen on these sightlines. In particular, they must also originate only in diffuse material on the Quintuplet sightline. 

Although an aromatic CH absorption band centered near 3.25~\mum\ has previously been detected towards young stellar objects embedded in dense cold molecular cloud cores, we demonstrate that its spectral profile and central wavelength differ significantly from those of the 3.3~\mum\ band observed toward Cyg OB2-12 and the Quintuplet. This indicates that the aromatic hydrocarbons in dense cloud cores are  different species than those seen in diffuse clouds.

Laboratory spectra of three carbonaceous dust analogs, produced through various deposition and processing techniques and intended to duplicate the interstellar 3.4~\mum\ and 6.85~\mum\ aliphatic bands, successfully reproduce the spectral profiles and relative band strengths for the aliphatic component of the dust observed toward Cygnus OB2-12 and the Quintuplet.

The detections on these sightlines of weak absorptions at 5.85 and 6.05~\mum, tentatively attributed to aliphatic carbonyl and quinone-like groups, suggests mild surface oxidation of carbon grains. 

Overall, the data support a structural model in which grains are composed of large, predominantly aromatic carbon cores coated with thin HAC mantles. This configuration yields an intermediate H/C ratio of $\simeq$0.13 and may also incorporate oxygen-bearing functional groups. The inferred structure is broadly aligned with the model proposed by \citet
{pendleton2002organic} in terms of the importance of the hydrogenated aliphatic carbon material, but extends it by incorporating new constraints from higher quality IR spectra that reveal that the HAC is a mantle on the aromatic carbon core, as postulated by \cite{chiar2013structure}.   

Looking ahead, theoretical models incorporating grain size distributions of carbonaceous cores with thin HAC mantles -- constrained by current and forthcoming observational data -- will help clarify the evolutionary pathways of dust in the diffuse ISM, as will laboratory experiments targeting a broader range of functional groups-especially those responsible for the mid-IR features. Additional observations of both diffuse and dense clouds will help reveal which dust components survive the transition from diffuse to dense environments, particularly at the boundaries of dense clouds where JWST can target background field stars. Observations of sources within the Cyg OB2 association, especially near Cyg OB2-12, will help determine whether the diffuse dust in this sightline has been altered by internal processes in the association, which contributes at least 4 of the total 10 magnitudes of extinction toward the star.

Although Cyg OB2-12 itself is too bright to be observed by JWST, the spectra presented here can be a useful point of comparison for interpreting recently published and future JWST observations of many additional diffuse interstellar sightlines.

\begin{acknowledgements}

YJP gratefully acknowledges support (GR107238) from the NASA Stratospheric Observatory for Infrared Astronomy (SOFIA) under award 75\_0107, which enabled an initial effort to obtain mid-IR data for Cyg OB2-12 using the EXES instrument. Although the early termination of the SOFIA mission limited the observing time and compromised both the completeness and quality of the data, the observations contributed to the verification of key features in the 6--7~\mum\ region. The award of SOFIA time enabled  related near-IR observations of Cyg OB2-12 to be obtained at the NASA Infrared Telescope Facility (IRTF), operated by the University of Hawai'i under contract with NASA, as part of a joint agreement between the IRTF and SOFIA. These near-IR data were essential to the present analysis. We also acknowledge the use of a reprocessed \textit{Spitzer Space Telescope} spectrum of Cyg OB2-12 first published in \citet{potapov2021dust}. Two of those authors (AP and JB) are coauthors on the present work; JB was responsible for the improved reduction of this archival data set which significantly improved its utility for this study, and AP provided one of the three laboratory analog datasets used here. Their contribution was instrumental in enabling the current analysis. YJP also thanks Dale P. Cruikshank for a  careful review of the manuscript, as those insights significantly improved clarity. TRG's research is supported by the international Gemini Observatory, a program of NSF NOIRLab, which is managed by the Association of Universities for Research in Astronomy (AURA) under a cooperative agreement with the U.S. National Science Foundation, on behalf of the Gemini partnership of Argentina, Brazil, Canada, Chile, the Republic of Korea, and the United States of America. AP acknowledges support from the Deutsche Forschungsgemeinschaft (Heisenberg grant PO 1542/7-1). BG acknowledges support of the Scientific and Technological Research Council of Türkiye -TÜBİTAK 2219. MEP acknowledges the grant INAF-Ricerca Fondamentale 2022. M.D. and S.Z. acknowledge support from the Research Fellowship Program of the European Space Agency (ESA).
 The first author (YJP) used the \textit{OpenAI GPT-4 model} (ChatGPT, \url{https://openai.com}) to a limited extent for assistance with LaTeX formatting and to improve the clarity of some sentences during the final stage of manuscript preparation. All scientific content, interpretations, and conclusions are solely those of the authors. We thank the anonymous referee for an insightful review, including constructive suggestions that improved the paper.
\end{acknowledgements}
\newpage
\bibliographystyle{aasjournal}
\bibliography{references.bib}

\end{document}